\documentclass[12pt,preprint]{emulateapj}

\shorttitle{Environmental dependence of galaxy colors}
\shortauthors{Wake et al.}

\begin{document}

\title{The environmental dependence of galaxy colors in intermediate redshift X-ray-selected clusters}

\author{David A. Wake$^{1,2,5}$, Chris A. Collins$^2$, Robert C. Nichol$^{1,5}$, Laurence R. Jones$^3$, and Douglas. J. Burke$^{2,4}$}
\altaffiltext{1}{Institute of Cosmology and Gravitation, University of Portsmouth, Mercantile House, Hampshire Terrace, Portsmouth, PO1 2EG, UK; david.wake@port.ac.uk}
\altaffiltext{2}{Astrophysics Research Institute, Liverpool John Moores University, Twelve Quays House, Egerton Wharf, Birkenhead CH41 1LD, UK}
\altaffiltext{3}{School of Physics and Astronomy, University of Birmingham, Birmingham B15 2TT, UK}
\altaffiltext{4}{Harvard Smithsonian Center for Astrophysics, 60 Garden Street, Cambridge, MA 02138, USA }
\altaffiltext{5}{Dept. of Physics, Carnegie Mellon University, 5000 Forbes Ave., Pittsburgh, PA 15217}

\begin{abstract}
We present a wide-field imaging study of the colors of bright galaxies ($<$ M$^*$+2) in a sample of 12 X-ray selected clusters and groups of galaxies at z $\sim$ 0.3.  The systems cover one of the largest ranges in X-ray luminosity (L$_X$ $\sim$ 10$^{43}$ - 10$^{45}$ erg s$^{-1}$), and hence mass, of any sample studied at this redshift. We find that the `red' galaxies form a tight color-magnitude relation (CMR) and that neither the slope nor zero-point of this relation changes significantly over the factor of 100 in X-ray luminosity covered by our sample. 
Using stellar population synthesis models we find our data allow a maximum possible change of 2 Gyrs in the typical age of the `red' galaxies on the CMR over the range of L$_X$ of our sample.
We also measure the fraction of blue galaxies (f$_b$) relative to the CMR in our clusters and find a low value of f$_b$ $\sim$ 0.1 consistent with other X-ray selected cluster samples. We find that there is no correlation between f$_b$ and L$_X$ over our large L$_X$ range.
However, we do find that both the CMR and f$_b$ do depend significantly on cluster radius, with the zero-point of the CMR shifting blueward in $B-R$ by 0.10 $\pm$ 0.036 magnitudes out to a radius of 0.75 times the virial radius. This color change is equivalent to a luminosity weighted age gradient of $\sim$ 2.5 Gyrs per log(radius) and is consistent with previous studies of the radial change in the zero-point of the CMR.
It thus appears that the global cluster environment, in the form of cluster mass (L$_X$), has little influence on the properties of the bright cluster galaxies, whereas the local environment, in the form of galaxy density (radius), has a strong effect. The range of $\sim$ 100 in L$_X$ corresponds to a factor of $\sim$ 40 in ram-pressure efficiency, thus suggesting that ram-pressure stripping, or other mechanisms that depend on cluster mass like tidal stripping or harassment, are unlikely to be solely responsible for changing the galaxy population from the `blue' star forming galaxies, that dominate low density environments, to the `red' passive galaxies that dominate cluster cores.

\end{abstract}


\section{Introduction}
\label{sec:intro}

The first clear evidence for galaxy evolution in clusters was presented by Butcher and Oemler (BO) in a series of papers in the late 70s and early 80s. Initially, they demonstrated that the fraction of blue galaxies (f$_b$) in two high redshift centrally concentrated clusters \citep{1978ApJ...219...18B} was significantly higher than that found in similar local clusters \citep{1978ApJ...226..559B}. This suggests that unlike the cores of local compact clusters which contain almost exclusively quiescent early-type galaxies, similar clusters at high redshift contain many star forming galaxies in addition to the early-types. This result was subject to significant scrutiny, particularly the required statistical removal of foreground and background galaxies \citep{1981ApJ...251..485M,1982ApJ...263..533D,1981ApJ...251L..75K}. In an attempt to resolve this, \citet{1984ApJ...285..426B} studied 33 galaxy clusters up to a redshift of 0.5. This showed an increasing trend of blue fraction with redshift, but with a large scatter. Several other studies have reported similar trends. For instance \citet{1995ApJ...439...47R} find an even faster increase, but do not limit their clusters to centrally concentrated ones, and \citet{2001ApJ...548L.143M} also confirm a similar trend using a large sample of Abell clusters.
  
Follow-up observations of the BO blue galaxies showed spectroscopic signatures of galaxies with either ongoing or recently-truncated star formation \citep{1987MNRAS.229..423C,1983ApJ...270....7D}, and morphological studies with HST have shown most of these galaxies to be late-type spirals but with an increased incidence of a disturbed or interacting appearance \citep{1994ApJ...430..107D,1997ApJ...474..561O,1994ApJ...430..121C,1998ApJ...497..188C}.  

However, it is not clear that the increased blue fraction is universal. \citet{1998MNRAS.293..124S}, using 10 bright X-ray clusters at z $\sim$0.25, have found no significant evidence for an increase in the blue fraction, and \citet{1993ApJ...404..521A} showed no increase in radio selected groups. It has also been suggested that the clusters studied at high redshift are not similar to those that were observed locally. For example, \citet{1999ApJ...516..647A} have shown that the Butcher and Oemler clusters have increasing X-ray luminosities with redshift.

Whilst evidence exists for strong evolution in the late-type population of clusters, the early-type galaxies have shown very little evidence of evolution beyond the passive ageing of their stellar populations. In particular, studies of the correlation between the colors and magnitudes of early-type galaxies \citep{1978ApJ...223..707S,1978ApJ...225..742S} in clusters \citep{1992MNRAS.254..589B,1992MNRAS.254..601B,2001MNRAS.326.1547T} have shown that this relation has a small scatter and evolves very little to high redshift \citep{1997ApJ...483..582E,1998ApJ...492..461S}. This slow evolution in the slope, zero-point and scatter of the color-magnitude relation (CMR) suggests that the stellar populations of early-type galaxies were formed at high redshift (z $\gtrsim$ 2), evolving purely passively since, with the slope being consistent with a metallicity sequence rather than an age sequence \citep{1997A&A...320...41K,1998A&A...334...99K}.


The only evolution in the early-type population in clusters appears to be in the number of S0 galaxies. A spectroscopic and HST study (MORPHS) of 10 rich galaxy clusters (0.37 $<$ z $<$ 0.56) found that the fraction of elliptical galaxies remained constant, but that the fraction of S0 galaxies decreased with redshift \citep{1997ApJ...490..577D}.
Further, they found an increase in both the star formation rate and in the number of galaxies showing evidence of recent star formation \citep{1999ApJ...518..576P} with redshift. However, a similar spectroscopic study of high redshift rich clusters, this time X-ray selected, found a very weak redshift dependence of the star formation rate with redshift \citep[$e.g.$ CNOC;][]{1999ApJ...527...54B}.

The decrease in S0 galaxies and the increase in blue star forming galaxies and galaxies with recently truncated star formation with redshift have often been linked. It has been suggested that spiral galaxies falling into the cluster experience either a burst of star formation followed by its rapid truncation or just a rapid truncation of their star formation. This results in the fading of the disk and the formation of an S0 galaxy. Several mechanisms that can occur in the cluster environment may be driving this transformation. These include ram-pressure stripping \citep{1972ApJ...176....1G,1998ApJ...509..587F,1999ApJ...516..619F,1999MNRAS.308..947A,2001ApJ...550..612F,2000Sci...288.1617Q}, galaxy harassment \citep{1996Natur.379..613M,1998ApJ...495..139M,1999MNRAS.304..465M}, tidal effects \citep{1990ApJ...350...89B,1996ApJ...459...82H,1998ApJ...509..587F}, and strangulation \citep{1980ApJ...237..692L,2000ApJ...540..113B,2001MNRAS.323..999D}. These processes have the affect of removing an infalling galaxies gas supply and thus halting ongoing star formation. 
This idea has been further reinforced by the change in galaxy properties with cluster radius. The Morphology-Radius/Density relation \citep{1974ApJ...194....1O,1980ApJ...236..351D} at low redshift is an excellent example of this and first inspired the idea that a galaxy's star formation may be halted by the cluster environment resulting in a morphological transformation \citep{1972ApJ...176....1G,1974ApJ...194....1O}. This relation was further shown to exist at high redshift \citep{1997ApJ...490..577D,2003ApJ...591...53T}. 

Similarly, \citet{1984ApJ...285..426B} showed that the blue fraction increased with radius in their clusters and this trend has generally been confirmed by other studies \citep[$e.g.$][]{1996ApJ...471..694A,2001MNRAS.321...18K,2002MNRAS.330..755F}. \citet{1997ApJ...488L..75B} show that the star formation rate depends strongly on cluster radius in the CNOC sample. In the same sample, they find no radial dependence of the fraction of post-star-burst/star-forming galaxies, unlike the MORPHS clusters \citep{1999ApJ...518..576P}. Both, however, conclude that the truncation of star formation is independent of the morphology-density relation. Recently, two large studies using the latest galaxy redshift surveys \citep[2dF,][SDSS]{2002MNRAS.334..673L,2003ApJ...584..210G} have found very clear relationships between star formation and density in the vicinity of clusters. Both studies find a characteristic density (radius) at which the average star formation rate rapidly drops.  

Some observational evidence also exists for an environmental dependence of the properties of elliptical galaxies in clusters. \citet{1996ApJ...471..694A} find that the galaxies on the CMR become bluer as a function of cluster radius in A2390 at a redshift of 0.23. \citet{2001MNRAS.326.1547T} find a similar trend in the Coma cluster, as does \citet{2002MNRAS.331..333P} in a combined sample of 11 rich clusters at z$\sim$0.1.

When trying to draw general conclusions regarding the evolution of cluster galaxies as discussed above, one particular concern is how cluster selection may have affected the results. It has already been noted that the BO sample has systematic differences as a function of redshift. A further concern comes from the fact that the clusters used were those that already had available data. This meant that they were often unusual objects \citep{1995deun.book.....S}, which was the reason they had been studied originally. This type of selection has often occurred \citep[$e.g.$ MORPHS;][]{1998ApJ...492..461S} and when other more careful selection criteria are used, differing results are found \citep[$e.g.$ CNOC;][]{1998MNRAS.293..124S}. It should also be noted that the clusters studied have mainly been massive rich clusters, as they provide the best statistics and are easiest to locate at high redshift. This may be of particular concern as they are very rare objects and are unlikely to be the evolutionary predecessors of the clusters with which they are compared locally \citep{1995MNRAS.274..153K}. Galaxy properties clearly depend on environment, and it may be that by studying only the most massive clusters a non-representative picture of galaxy evolution has been generated.

This paper presents further results from a comprehensive study of galaxy populations in X-ray selected clusters at both intermediate and high redshift. The clusters were selected to cover a wide range in X-ray luminosity (L$_X$) and all were imaged over a wide field in three bands. The wide range in X-ray luminosity and hence mass and the wide-field imaging allows us to simultaneously investigate how the cluster galaxy population depends on both cluster mass and cluster radius. In \citet[][Paper I]{2002MNRAS.330..755F} we presented measurements of the blue fraction in an initial sample of 8 clusters over a large redshift range.
In this paper 12 intermediate redshift clusters from the sample are used to investigate how the blue fraction and CMR depend on both X-ray luminosity and cluster radius.

Throughout this paper a cosmology with $\Omega_M$ = 0.3, $\Omega_{\Lambda}$ = 0.7 and $H_0$ = 70 km s$^{-1}$ Mpc$^{-1}$ is assumed.

\section{Cluster Selection and Data}

\begin{table*}
  \begin{center}
    \caption{Details of the cluster sample\label{tab:clusamp}}
    \begin{tabular}{l  l  c  c  c  c  c  l} 
      \multicolumn{1}{l}{Cluster} &
      \multicolumn{1}{l}{Catalogue} &
      \multicolumn{1}{l}{RA} &
      \multicolumn{1}{l}{Dec} &
      \multicolumn{1}{c}{Redshift} &
      \multicolumn{1}{c}{L$_X$} &
      \multicolumn{1}{c}{L$_X$ error} &
      \multicolumn{1}{l}{Telescope}\\%
      \hline \hline 
      RX1633.6+5714 & WARPS   & 248.4254 & 57.2366  & 0.239 &  0.10 & 0.02 & INT \\
      RX0333.0-3914 & SSHARC  & 53.2755  & -39.2305 & 0.245 &  0.12 & 0.02 & CTIO\\
      RX0210.4-3929 & WARPS   & 32.6112  & -39.4908 & 0.273 &  0.16 & 0.02 & CTIO\\
      RX0054.0-2824 & WARPS   & 13.5150  & -28.4039 & 0.293 &  0.24 & 0.03 & CTIO\\
      RX1606.7+2329 & WARPS   & 241.6825 & 23.4875  & 0.310 &  0.69 & 0.06 & INT\\
      RX2237.0-1516 & BSHARC  & 339.2504 & -15.2766 & 0.299 &  1.09 & 0.08 & CTIO\\
      MS0407.2-7123 & EMSS    & 61.6908  & -71.2750 & 0.229 &  1.87 & 0.41 & CTIO\\
      RX1418.5+2510 & WARPS   & 214.6279 & 25.1813  & 0.294 &  2.96 & 0.20 & INT \\
      RX0256.5+0006 & BSHARC  & 44.1370  & 0.1032   & 0.360 &  3.44 & 0.17 & CTIO \\
      MS0353.6-3642 & EMSS    & 58.8733  & -36.5641 & 0.320 &  7.05 & 1.38 & CTIO \\
      MS1455.0+2232 & EMSS    & 224.3129 & 22.3427  & 0.259 &  12.81 & 0.64 & INT\\
      MS2137.3-2353 & EMSS    & 325.0633 & -23.6608 & 0.313 &  15.48 & 2.11 & CTIO\\
      \hline
    \end{tabular}
    \end{center}
	{Details of the cluster sample, catalogue, coordinates of the center, redshift, X-ray luminosity, error in the X-ray luminosity and Telescope. L$_X$ is given for the ROSAT band (0.5 - 2.0 keV) in units of 10$^{44}$ erg s$^{-1}$.}
\end{table*}

The 12 clusters presented here are selected to cover a large range of X-ray luminosities and hence cluster masses over a small redshift interval at z $\sim$ 0.3. Because they are all selected using their X-ray emission and are at similar redshifts, comparison of their galaxy populations should be free of any selection effects driven by the cluster selection technique. 
The clusters were drawn from 4 X-ray selected cluster samples: the Extended Medium Sensitivity Survey \citep[EMSS;][]{1990ApJS...72..567G}, the Southern Serendipitous High-redshift Archival ROSAT Cluster \citep[SSHARC;][]{1997ApJ...479L.117C,2003MNRAS.341.1093B}, the Bright Serendipitous High-redshift Archival ROSAT Cluster \citep[BSHARC][]{1999ApJ...521L..21N,2000ApJS..126..209R} and the Wide Angle ROSAT Pointed Survey \citep[WARPS;][]{1997ApJ...477...79S,2002ApJS..140..265P}. The low flux limits of SHARC and WARPS allow the selection of clusters with low L$_X$, whereas the large area of the EMSS allows the selection of rarer high L$_X$ clusters. Table \ref{tab:clusamp} lists the name, original catalog, coordinates of the center, redshift, X-ray luminosity, and telescope (with which the observations were made) of the cluster sample.

\subsection{Observations}
Each cluster was imaged over a wide field of view through three filters, $B$, $V$ and $R$, in the Johnson-Cousins system. These filters correspond approximately to $U$, $B$, and $V$ at rest at the cluster redshift, minimizing any K-corrections required.
The imaging was undertaken using two telescopes: the Issac Newton Telescope (INT) at the Roque de Los Muchachos Observatory on the Island of La Palma and the Blanco Telescope at the  Cerro Tololo Inter-American Observatory (CTIO), Chile.
Each telescope was equipped with a wide-field mosaic CCD camera. The INT Wide Field Camera \citep{Ives1996} consists of 4 thinned EEV 2k$\times$4k CCDs mounted at the INT prime focus in an `L' configuration. Each CCD has a pixel size of 13.5 $\mu$m corresponding to 0.33 arcseconds on the sky. There are 2048 by 4100 pixels in the useful imaging area of each CCD, giving a field of view of 22.8 by 11.4 arcminutes. The 4 CCDs together give a total imaging area of $\sim$ 0.29 deg$^2$. The CTIO MOSAIC II camera consists of 8 2048 by 4096 SITe CCDs mounted in 2 rows of four. Each pixel is 15 $\mu$m corresponding to 0.27 arcseconds on the sky, giving each CCD dimensions of 9.2 by 18.4 arcminutes, and a total imaging area of $\sim$ 0.38 deg$^2$. The gaps between the CCDs are $\sim$ 40 pixels or 11 arcseconds.

\begin{table}
  \begin{center}
    \caption{Description of the INT observations.\label{tab:INTobs}}
    \begin{tabular}{c  c  c  c  c c} 
      \multicolumn{1}{c}{Cluster} &
      \multicolumn{1}{c}{Filter} &
      \multicolumn{1}{c}{Night} &
      \multicolumn{1}{c}{Exposure Time(s)} &
      \multicolumn{1}{c}{Number} &
      \multicolumn{1}{c}{Seeing}\\
      \hline \hline 
MS1455.0 &  B  &  2  &  2250  &  3  &  1.22\\
	 &  V  &  1  &  2400  &  2  &  1.58\\
	 &  R  &  1  &  2700  &  3  &  1.56  \\
RX1418.5 &  B  &  3  &  3600  &  3  &  1.23\\
	 &  V  &  3  &  1200  &  1  &  0.93\\
	 &  R  &  3  &  1200  &  1  &  0.87\\
RX1633.6 &  B  &  3  &  1200  &  1  &  1.15\\ 
	 &  B  &  4  &  1800  &  2  &  1.20\\
	 &  V  &  4  &  1200  &  1  &  0.88\\
	 &  R  &  4  &  1200  &  1  &  0.79\\
RX1606.7 &  B  &  4  &  3600  &  3  &  1.11\\
	 &  V  &  4  &  1200  &  1  &  1.11 \\ 
	 &  R  &  4  &  1200  &  1  &  0.93\\
      \hline
    \end{tabular}
  \end{center}
    {Summary of the cluster observations made at the INT from 17$^{th}$-20$^{th}$ June 1999. For each cluster the filter, the night, the total exposure time, the number of exposures and the mean seeing are listed.}
\end{table}

\begin{table}
  \begin{center}
    \caption{Description of the CTIO observations.\label{tab:CTIOobs}}
    \begin{tabular}{c  c  c  c  c c} 
      \multicolumn{1}{c}{Cluster} &
      \multicolumn{1}{c}{Filter} &
      \multicolumn{1}{c}{Night} &
      \multicolumn{1}{c}{Exposure Time(s)} &
      \multicolumn{1}{c}{Number} &
      \multicolumn{1}{c}{Seeing}\\
      \hline \hline 
RX0054.0	&   B   &   1   &  1650	&  3   &   1.23\\
		&   V   &   1   &  900	&  3   &   1.12\\
		&   R   &   1   &  630	&  3   &   1.17\\
RX0210.4	&   B   &   1   &  1650	&  3   &   1.26\\
		&   V   &   1   &  900	&  3   &   1.25\\
		&   R   &   1   &  630	&  3   &   1.26\\
RX0256.5   	&   B   &   2   &  1650	&  3   &   1.70\\
		&   V   &   1   &  900	&  3   &   0.87\\
		&   R   &   1   &  630	&  3   &   0.84\\
RX0333.0	&   B   &   1   &  1650	&  3   &   0.99\\
		&   V   &   1   &  900	&  3   &   0.86\\
		&   R   &   1   &  630	&  3   &   0.92\\
MS0353.6  	&   B   &   2   &  1650	&  3   &   1.67\\
		&   V   &   2   &  900	&  3   &   1.49\\
		&   R   &   2   &  630	&  3   &   1.49\\
MS0407.2	&   B   &   2   &  1500	&  3   &   1.95\\
		&   V   &   2   &  750	&  3   &   2.06\\
		&   R   &   2   &  630	&  3   &   1.84\\
MS2137.3	&   B   &   1   &  1650	&  3   &   1.41\\
		&   V   &   1   &  900	&  3   &   1.21\\
		&   R   &   1   &  630	&  3   &   1.13\\
RX2237.0	&   B   &   1   &  1650	&  3   &   1.20\\
		&   V   &   1   &  900	&  3   &   1.07\\
		&   R   &   1   &  630	&  3   &   1.05\\
      \hline					   
    \end{tabular}				   
  \end{center}
    {Summary of the cluster observations made at the CTIO from 26$^{th}$-27$^{th}$ September 2000. For each cluster the filter, the night, the total exposure time, the number of exposures and the mean seeing are listed.}
\end{table}

The INT data were acquired on 4 nights from 17-20th June 1999. The observing conditions were generally good through the 4 nights, except for the first night during which light cirrus cloud was present. Despite this night being non-photometric, observations of clusters were still made using slightly longer exposure times with calibration images of the same fields taken on the following night. It was also noted that Saharan dust was present in the upper atmosphere throughout the observations. This is a common feature at the Roque de Los Muchachos Observatory and is known to show `gray' absorption, increasing the atmospheric extinction equally over all wavelengths \citep{INTtn31}. The CTIO data were taken on the nights of the 26th and 27th of September 2000. The observing conditions were good on both nights with no cloud cover. Tables \ref{tab:INTobs} and \ref{tab:CTIOobs} give details of the observations.

\subsection{Data Reduction}

The basic reduction steps were undertaken using the IRAF package MSCRED \citep{1998adass...7...53V}. This package was written specifically to reduce data from the two NOAO mosaic cameras, one of which is the MOSAIC II camera on the CTIO Blanco telescope. The main advantage of this over other data reduction packages is its ability to operate on the multi-CCD data simultaneously, without the need to split each image frame into separate CCD images and process them individually. It was also possible to use this package with non-NOAO Mosaic data such as from the INT WFC. Before this could occur, modifications to the headers of the INT images describing the relative position of the CCDs to each other were required.

Both data sets were reduced in a similar manner. Only when a different procedure was applied to each data set will the individual data sets be identified. 

In general the data reduction proceeded in the standard manner for CCD imaging. After bias subtraction the images were flat-fielded using dome flats. The R band data from the INT had to be corrected for fringing. A fringe frame was generated from the object frames and then subtracted from the individual frames. Sky flats were then generated by median combining the object frames, and a further flat-field correction applied using these. Typically this produced frames with a maximum variation in the sky level of $\lesssim$ 0.5\% and $\lesssim$ 1\% for the CTIO and INT images respectively. Cosmic rays, saturated pixels and stellar bleed trails were identified and added to the bad pixels masks. In addition to these steps a linearity correction was required for three of the INT WFC CCDs. This non-linearity was investigated previously by the Cambridge Astronomical Survey Wide Field Survey team \citep{Mcmahon1999} and found to be constant with time. They characterized the non-linearity and have provided equations to convert the measured values of CCDs 2, 3 and 4 to the value of CCD 1. A further correction was also required for the CTIO MOSAIC camera data to remove the 'cross-talk' in the controller electronics. This can cause pixel values for one amplifier to be affected by another and induce `ghost images'. This is well understood for the CTIO MOSAIC camera and a model has been developed to correct the data. 

Wide-field imaging requires a step not normally necessary in CCD data reduction. The curvature of the focal plane causes a significant variation to the pixel size across the CCD array. After flat-fielding all pixels have the same sky level which is incorrect for pixels of different sizes and leads to a varying photometric zero-point. To correct for this an accurate astrometric solution was determined for each image, which was then used to rebin the image to a constant pixel scale. Before rebining, the pixel scale changed by $\sim 4\%$ in the INT WFC and $\sim 8\%$ in the CTIO MOSAIC II camera from the center to the edge of the array and would thus cause a significant error in the photometry across the field.

Before combining the individual images into final stacked images of each cluster in each band, the photometric scales were matched, correcting for variations in the sky level, extinction and exposure times. This was done by comparing the flux of a number of stars that appear on each image.
The corrected images are then stacked using a mean, excluding pixels in the bad pixel mask. 

The photometric calibration was achieved using images of standard star fields \citep{1992AJ....104..340L} taken throughout each night. Fields covering a large area were chosen so that it was possible to place standard stars on every CCD with one pointing. For the CTIO data there were sufficient standards to fit for a zero-point, an extinction and a color correction term on each CCD on each night. As would be expected the extinction terms were all equivalent within the errors and so a weighted mean was taken. The zero-point and color terms were then refitted. Although the color terms may be expected to vary slightly between CCDs the small size of these corrections meant that they were equivalent within their errors and a weighted mean was again used to give a final color term for each CCD. The zero-points were then refitted using the mean extinction and color terms. Again they were found to be sufficiently consistent (rms $<$ 0.01) to combine and use as one term for the whole mosaic. Although using one photometric solution for the whole array may slightly increase the photometric errors it greatly simplifies the photometric measurements for the stacked images since spatial offsets between the individual images mean that a single patch of the stacked image can have contributions from several CCDs.

The INT calibration proceed in a similar manner. However, fewer images of standard fields were observed over too low a range in airmass preventing an accurate determination of the extinction. Instead, extinction coefficients were used from the Carlsberg Meridian Telescope a photometric monitoring telescope on the observatory site. As for the CTIO, data zero-points and color terms were first determined for each individual CCD to ensure they were consistent and then combined to determine zero-points for the whole array. 

\subsection{Constructing Galaxy Catalogues}

Object detection and photometry were performed using the Source Extractor package \citep[SExtractor;][]{1996A&AS..117..393B}. Objects were detected on the R band images only, with photometry then being performed on all three images using the objects from the R detection. A Gaussian filter of size equal to the image seeing was used during the detection process to better differentiate between faint sources and noise peaks. The areas around very bright sources were also masked as these can effect SExtractor's background determination leading to poor photometry. Both total photometry, using a modified Kron aperture \citep{1980ApJS...43..305K}, and aperture photometry were performed. 

It is known that SExtractor's total photometry estimate suffers from a decrease in the included flux with decreasing total flux. In order to calculate a correction to the total magnitude, isolated galaxies were identified on each image. Using the IRAF task PHOT, the magnitudes of these isolated galaxies within a large aperture were calculated in order to estimate the true total magnitude.
 For bright objects, the difference between the Sextractor and true total magnitudes was approximately constant at $\sim$ 0.04 magnitudes or $\sim$ 4$\%$. This rose to $\sim$ 0.2 magnitudes or $\sim$ 15$\%$ for the faintest objects. These corrections were applied to all the total magnitudes. Apertures with a physical diameter of 20 Kpc were used to calculate the colors, representing a minimum size of at least twice the seeing, thus ensuring that the small seeing variations between images do not effect the colors.
The observed colors of each object were used to apply the color corrections determined from the standard solutions giving the final magnitudes for each object. In order to remove any spurious detections, objects that were not detected in at least two bands were removed from the catalogue. Finally, the number of bad pixels present in an object was calculated from the exposure masks and added to the catalogue. Any object with more than 10$\%$ bad pixels in a band had the magnitude in that band flagged. The final catalogues of all measured objects in a field contained between 15433 and 39977 objects.

\subsection{Star-Galaxy Separation}

\begin{figure}
\includegraphics[angle=-90,scale=.70]{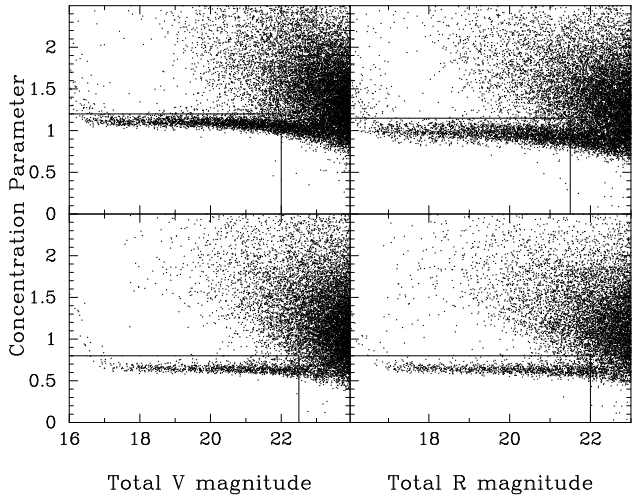}
\caption{\label{fig:conc} The concentration parameter as a function of magnitude in the $V$ and $R$ filters. Top panels show cluster MS2137.3 and the bottom show RX0256.5. The lines show the division selected to divide stars from galaxies.}
\end{figure}

\begin{figure*}
\plotone{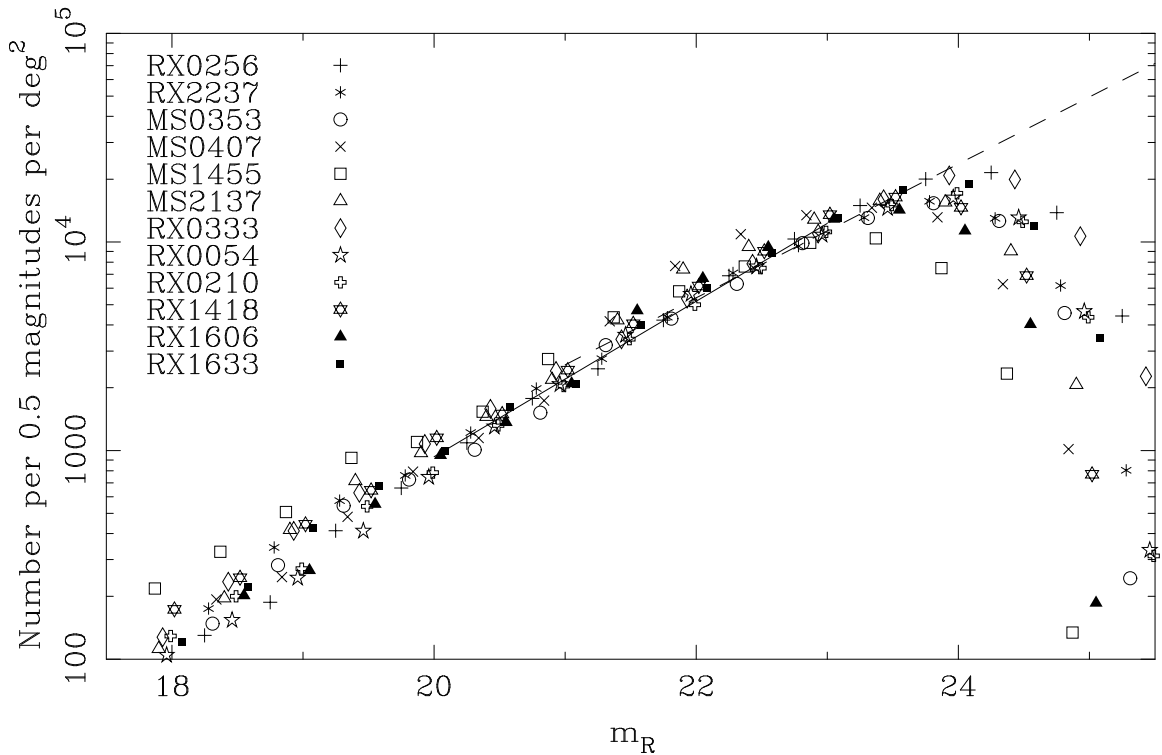}
\caption{\label{fig:NummagR}The $R$ band number density-magnitude relation for all 12 cluster fields. The solid line shows the fitted relation of \citet{1991MNRAS.249..498M}. The dashed line shows the relation of \citet{1995ApJ...449L.105S}.}
\end{figure*}

In order to differentiate stars from galaxies a method similar to that presented in \citet{1991MNRAS.249..498M} was used. A concentration parameter was defined as the magnitude within 4 pixels divided by the total magnitude. 
Figure \ref{fig:conc} shows the concentration parameter as a function of magnitude for two example cluster fields. The stars are visible as the horizontal linear feature in the plots and are clearly separated from the galaxies at bright magnitudes. The solid lines define the region chosen to represent the stars in each plot. These criteria had to be selected for each cluster individually as this parameter is dependant on the image seeing, with increasing seeing moving the locus of stars to higher values and greater width. The seeing was observed to vary across the wide-field of our images, up to 10$\%$ in the CTIO images and 20$\%$ in the INT images. The effect of this radial variation in the seeing is to just broaden the stellar locus, reducing the magnitude with which the separation can occur, but not causing any systematic change in the separation with radius.  The $B$-band data were not used for the final separation as they displayed a larger scatter in the stellar locus. The separation was typically effective to an $R$ magnitude of $\sim$ 22. 
Fainter than this, the number counts are dominated by galaxies. Any remaining stars will be included in the field region in equal proportions as in the cluster regions and thus will be removed by the statistical background subtraction (see Section \ref{sec:fcorr}).

To confirm the reliability of this method, the star and galaxy classifications in the field of RX0256.5 were compared to those of the Sloan Digital Sky Survey \citep[SDSS;][]{2000AJ....120.1579Y} which also images this field. The SDSS star-galaxy separation is reliable to m$_{R} \simeq$ 22, comparable to the magnitude at which the separation was possible here. Of the 2558 galaxies, 52 were classified as stars by the SDSS, and of the 1155 stars, 87 were classified as galaxies by SDSS. The SDSS star-galaxy separation is approximately 95$\%$ accurate at these depths \citep{2001adass..10..269L}, suggesting that the separation here is at least as accurate.

The final step in the construction of the galaxy catalogues is to correct for galactic extinction. To apply the required corrections the maps of \citet{1998ApJ...500..525S} were used. Typically, one extinction correction is applied per field. However, the resolution of these maps is sufficient to map variations over the wide fields of the images used here. Investigation showed that the extinction could vary by up to 25$\%$ across a field. Therefore, the extinction was calculated for each galaxy individually by using the interpolation options available in the codes provided with the \citet{1998ApJ...500..525S} maps.

\subsection{Photometric Checks and Completeness}
\label{sec:photchk}

Before using the photometry for our scientific investigation it is reassuring to check its accuracy against external sources and to determine if the observations have achieved the desired depth and accuracy. Fortunately, two of the fields imaged here have existing photometry from other sources that may be used to check the reliability of the photometry presented here. 

\citet{1998AJ....116..102T} measured Johnson $B$ and $B - R$ photometry for 99 stellar objects in the field of RX1418.5. Sixty-five of these objects have photometry measured by SExtractor, of which 41 are stars. Comparing with our B and R photometry for these 41 stars shows a median difference of less than 0.03 and a $rms$ $\lesssim$ 0.1 showing a good consistency between the data sets to m$_B$ = 22.


The field of RX0256.5 has photometry available from the Sloan Digital Sky Survey in the Sloan filters ($u,g,r,i,z$). Using the conversions from \citet{2002AJ....123.2121S} and the Sloan PSF magnitudes, the equivalent total $B$, $V$, and $R$ magnitudes were calculated for all the stars present in both data sets. The conversions were found to be less reliable at either end of the $g-r$ colour range and so the sample was restricted to 0.5 $\le g-r \le$ 1.2 leaving 1076 stars in common. The median offset was found to be $<$ 0.03 in all of the filters with an $rms$ $<$ 0.09 to m$_R$ = 22, showing excellent agreement between the two data sets.


Figure \ref{fig:NummagR} shows the number magnitude relations (NMR) in the $R$-band using the SExtractor total magnitudes. Reassuringly, the relations are similar for all the cluster fields but do show some scatter as would be expected from cosmic variance. The mean density can vary by almost a factor of two between cluster fields. This illustrates the importance of individual estimates of the background density when correcting cluster galaxy counts (see Section \ref{sec:fcorr}). Plotted on the R-band NMR are the relations derived by \citet{1991MNRAS.249..498M} and \citet{1995ApJ...449L.105S}. These are consistent with the NMRs of the cluster fields providing another confirmation of the photometry. 

These relations can be used to estimate the completeness of the galaxy catalogues. The NMRs clearly begin to turn over as the completeness declines. Table \ref{tab:cmfitB-R} lists the estimated completeness in the detection bands and the original target equivalent to M$_V$ = -19 for an elliptical galaxy at redshift zero. These values were estimated by observing the position at which the NMR appears to turn over. This level of precision was adequate, as the questions considered in this paper typically use galaxies at least $\sim$ 1 magnitude brighter than these limits.

\section{Analysis and Results}
\subsection{Defining a Characteristic Radius}

The variation in the mass of our cluster sample means that each cluster has a different physical size. In order to compare the galaxy populations between clusters we must define some characteristic radius that is equivalent for all the clusters. If we just use a fixed physical scale it may include just the core of the most massive clusters and substantially beyond the core of the least massive.
To this end \citet{1978ApJ...226..559B} defined a characteristic radius, R$_{30}$, within which they measured the blue fraction. R$_{30}$ is defined as the radius containing 30\% of the total number of galaxies in a cluster. To calculate this, it is necessary to integrate the cluster density profile up to the edge of the cluster (R$_{100}$) after subtracting the background galaxy density. Most other studies of the Butcher-Oemler effect have attempted to calculate R$_{30}$ for their clusters \citep[$e.g.$][]{1998MNRAS.293..124S,2001MNRAS.321...18K,2002MNRAS.330..755F,2002MNRAS.331..333P}. Apart from the original Butcher and Oemler papers, details of the calculation of R$_{30}$ are scarce, particularly of how R$_{100}$ and the background level were determined. 

\begin{table*}
  \begin{center}
    \caption{Cluster X-ray properties.\label{tab:Virialrad}}
    \begin{tabular}{l  c  c  c  c  c  c} 
      \multicolumn{1}{l}{Cluster} &
      \multicolumn{1}{c}{Redshift} &
      \multicolumn{1}{c}{L$_X$ ROSAT} &
      \multicolumn{1}{c}{L$_{bol}$} &
      \multicolumn{1}{c}{$k$T} &
      \multicolumn{1}{c}{$R_V$} &
      \multicolumn{1}{c}{$R_V$}\\ 
      \multicolumn{1}{l}{} &
      \multicolumn{1}{c}{} &
      \multicolumn{1}{c}{$10^{44}$ erg s$^{-1}$} &
      \multicolumn{1}{c}{$10^{44}$ erg s$^{-1}$} &
      \multicolumn{1}{c}{keV} &
      \multicolumn{1}{c}{Mpc} &
      \multicolumn{1}{c}{arcmins}\\ 
 
      \hline \hline 
      RX1633.6 &  0.239 &  0.10 $\pm$ 0.02 &  0.24 $\pm$ 0.05 &  1.53 $\pm$ 0.27 &  1.17 $\pm$ 0.10 &  5.16 $\pm$ 0.44  \\
      RX0333.0 &  0.245 &  0.12 $\pm$ 0.02 &  0.30 $\pm$ 0.06 &  1.67 $\pm$ 0.27 &  1.22 $\pm$ 0.11 &  5.27 $\pm$ 0.43 \\
      RX0210.4 &  0.273 &  0.16 $\pm$ 0.02 &  0.41 $\pm$ 0.06 &  1.86 $\pm$ 0.28 &  1.26 $\pm$ 0.10 &  5.02 $\pm$ 0.40  \\
      RX0054.0 &  0.293 &  0.24 $\pm$ 0.03 &  0.66 $\pm$ 0.10 &  2.19 $\pm$ 0.31 &  1.34 $\pm$ 0.10 &  5.10 $\pm$ 0.38  \\
      RX1606.7 &  0.310 &  0.69 $\pm$ 0.06 &  2.24 $\pm$ 0.24 &  3.47 $\pm$ 0.33 &  1.66 $\pm$ 0.08 &  6.08 $\pm$ 0.29  \\
      RX2237.0 &  0.299 &  1.09 $\pm$ 0.08 &  3.85 $\pm$ 0.35 &  4.28 $\pm$ 0.31 &  1.86 $\pm$ 0.07 &  6.98 $\pm$ 0.26  \\
      MS0407.2 &  0.229 &  1.87 $\pm$ 0.41 &  3.63 $\pm$ 0.95 &  4.32 $\pm$ 0.52 &  1.98 $\pm$ 0.12 &  9.03 $\pm$ 0.54  \\
      RX1418.5 &  0.294 &  2.96 $\pm$ 0.20 & 12.47 $\pm$ 1.10 &  6.69 $\pm$ 0.45 &  2.34 $\pm$ 0.08 &  8.87 $\pm$ 0.30  \\
      RX0256.5 &  0.360 &  3.44 $\pm$ 0.17 & 14.71 $\pm$ 1.01 &  6.93 $\pm$ 0.48 &  2.25 $\pm$ 0.08 &  7.46 $\pm$ 0.27 \\
      MS0353.6 &  0.320 &  7.05 $\pm$ 1.38 & 17.53 $\pm$ 4.31 &  8.13 $\pm$ 0.36 &  2.52 $\pm$ 0.06 &  9.03 $\pm$ 0.22 \\
      MS1455.0 &  0.259 & 12.81 $\pm$ 0.40 & 36.74 $\pm$ 1.69 &  5.60 $\pm$ 0.29 &  2.20 $\pm$ 0.06 &  9.15 $\pm$ 0.25 \\
      MS2137.3 &  0.313 & 15.48 $\pm$ 2.11 & 45.65 $\pm$ 8.42 &  5.00 $\pm$ 1.10 &  1.99 $\pm$ 0.22 &  7.23 $\pm$ 0.80 \\
      \hline
    \end{tabular}
  \end{center}
    { The calculated bolometric luminosities, temperatures and virial radii for all clusters using the L$_X$-T relation of \citet{2002ApJ...578L.107V}. For MS0353.6, MS1455.0 and MS2137.3 the temperatures quoted are those taken from \citet{1997ApJ...482L..13M} and \citet{1999MNRAS.305..834E}. The errors in $R_V$ are calculated from the L$_X$ error along with the errors on the L$_X$-T relation.}
\end{table*}

Here we attempted to determine R$_{30}$ by the following procedure. For each cluster, the density profile was calculated in 0.25 arcminute bins, by counting all galaxies whose centers lay in each bin up to a limiting $R$ magnitude of 22.5. The area in each radial bin was then calculated using the exposure masks of each image. Only the areas which had maximum exposure in all three bands and the galaxies within those areas were included. A further correction to the area was required to take into account the effects of large objects and crowding. Large objects potentially obscure galaxies in front of or behind them. This can be a problem, particularly in the cores of dense clusters where many bright ellipticals lie. For all objects larger than 900 pixels, an estimate was made of the area brighter than the deblending threshold. These estimates were then summed and subtracted from the previously calculated area.

\begin{figure}
\plotone{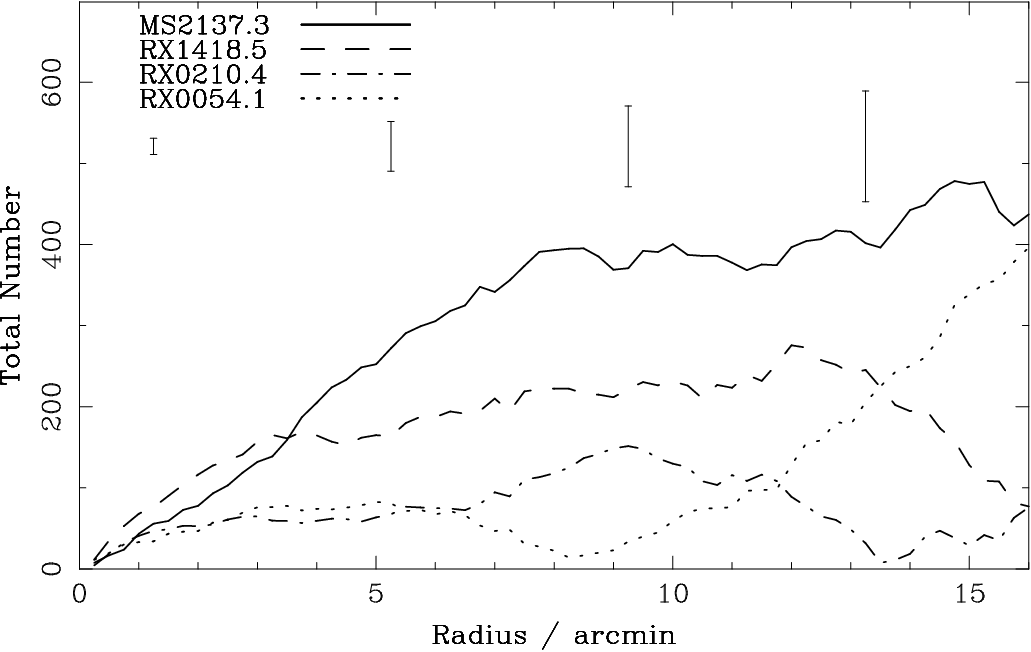}
\caption{\label{fig:denrad1}The total number of galaxies as a function of radius from each cluster's center, for a representative sub-sample of our clusters. Typical error bars are shown above the lines.}
\end{figure}

Figure \ref{fig:denrad1} shows the integrated number of galaxies as a function of radius for four example clusters. The integrated number was calculated by subtracting a background density typically calculated as the mean density in the region 11 - 17 arcminutes from the cluster center. In some instances, this was adjusted to avoid obvious over-densities in the background. 

In order to calculate R$_{30}$, it was necessary to first determine the cluster `edge', R$_{100}$. This was difficult, as the edge was lost in the variations of the background density. In most clusters, a number of radii seemed applicable as a possible R$_{100}$ and it was unclear which, if either, represents the `real' cluster edge.

\begin{figure}
\plotone{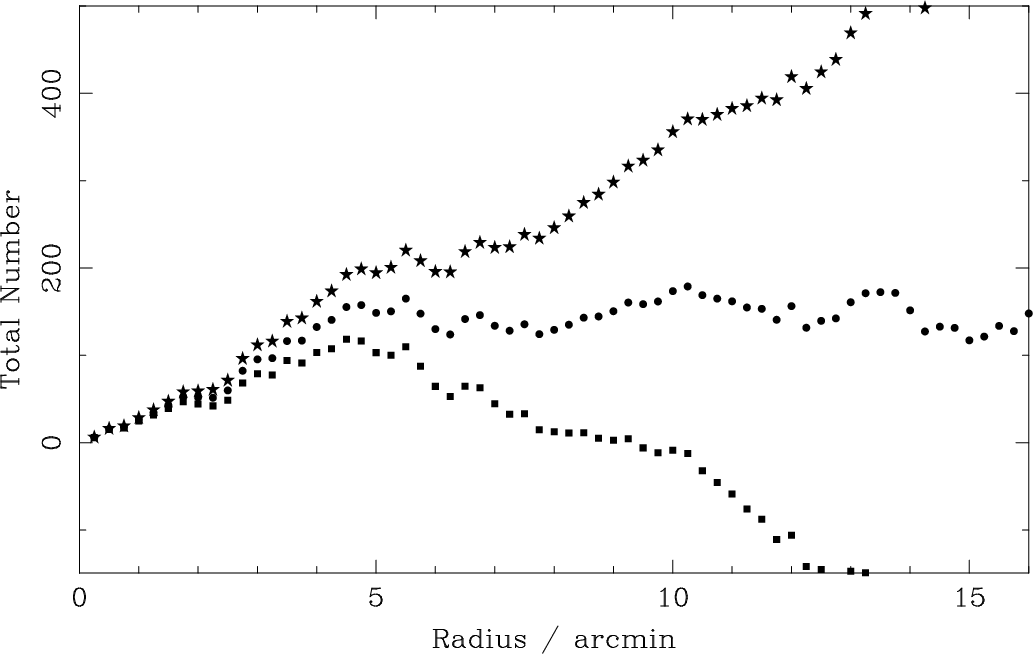}
\caption{\label{fig:denradbck}The effect of varying the background density on the total number radius relations of MS1455.0. The relations shown are calculated with the background density calculated as the mean density in the region 11 - 17 arcminutes from the cluster center with that density plus (stars) and minus (squares) its $rms$ error.}
\end{figure}

The profiles in Figure \ref{fig:denrad1} illustrate the difficulty in calculating R$_{30}$ particularly for the low L$_X$ clusters. Aside from determining the cluster edge, the choice of background density can also have a large effect on R$_{30}$. The unphysical nature of the very rapidly rising or falling total number-radius plots for some clusters ($e.g.$ RX0054.1) demonstrates this sensitivity. To further highlight this problem, Figure \ref{fig:denradbck} shows how the total number of galaxies as a function of radius varies when the background density is changed. The background level is changed by its $\pm$ 1 $\sigma$ error, resulting in wild variations in the total number profiles. This may be a particular problem for previous studies that have used a small field of view to image the clusters and then a separate global background estimate. The difficulty in determining R$_{30}$ is further illustrated using the cluster MS1455.0. \citet{1998MNRAS.293..124S} calculated an R$_{30}$ of 2.2 arcmins for this cluster, whereas \citet{2002MNRAS.330..755F} found 3.75 arcmins, and this study either 1.56 or 3.07 arcmins depending on the choice of the cluster edge. This could potentially be a significant problem for blue fraction studies, particularly when comparing samples from different authors, considering that the blue fraction has been shown to depend on radius \citep{1984ApJ...285..426B,1996ApJ...471..694A,2001MNRAS.321...18K,2001ApJ...547..609E}.

In an attempt to reduce some of the systematic uncertainties in calculating R$_{30}$, it was decided to define another characteristic radius using the X-ray properties of this cluster sample. Assuming the cluster core is virialised, in a simple self similar model \citep{1998ApJ...495...80B,1998ApJ...503..569E} the virial radius, $R_V$, is given by  

\begin{equation}
\label{eq:Rv}
R_V=3.80\beta_T^{-1/2}\Delta_z^{-1/2}(1 + z)^{-3/2}\left(\frac{kT}{10 keV}\right)^{1/2}h_{50}^{-1},
\end{equation}
with
\begin{equation}
\label{eq:dz}
\Delta_z = (\Delta_c(\Omega_z,\Lambda)\Omega_0)/(18\pi^2\Omega_z)
\end{equation}

taken from \citet{2002A&A...389....1A},
where $\Delta_c(\Omega_z,\Lambda)$ is the density contrast, and $\beta_T$ is the normalization of the virial relation $GM_V/2R_V = \beta_TkT$. A value of $\beta_T$ = 1.05 \citep{1996ApJ...469..494E} is used here. An analytic approximation for $\Delta_c(\Omega_z,\Lambda)$ is given in \citet{1998ApJ...495...80B}.

Only three clusters in our sample have measured temperatures to use in the above relation: MS0353.6, MS1455.0 \citep{1997ApJ...482L..13M}, and MS2137.3 \citep{1999MNRAS.305..834E}.
For the rest of the sample, it was necessary to convert the measured L$_X$ to temperature.  This was done using the redshift dependent $L_X-T$ relation,

\begin{equation}
\label{eq:LT}
T =  \left(\frac{L_{bol}}{B(1 + z)^A}\right)^{1/\alpha}
\end{equation}
where B = 3.11 $\pm$ 0.27 $\times$ 10$^{44}$ h$^{-2}$ erg s$^{-1}$, A = 1.5 $\pm$ 0.3, $\alpha$ = 2.64 $\pm$ 0.27 \citep{1998ApJ...504...27M,2002ApJ...578L.107V}, and L$_{bol}$ is the bolometric luminosity.

A conversion from the ROSAT L$_X$ to L$_{bol}$ was required before the above $L_X-T$ relation could be used. This conversion was calculated using a model X-ray spectrum. The shape of the spectrum is dependent on the temperature, and hence so is the conversion. The two steps were therefore combined iteratively, with an initial estimate of temperature used to calculate L$_{bol}$, which was in turn used in the $L_X-T$ relation to estimate the temperature, which was again fed into the L$_{bol}$ conversion and so on until convergence. This procedure converged rapidly within 3-5 iterations. 

The complexity of the above method, combined with the existence of significant scatter in the $L_X-T$ relation, and the requirement of a virialised cluster core, call into question the reliability of $R_V$ and its superiority over R$_{30}$. Since all of these clusters show reasonably strong X-ray emission, the likelihood of the cores being virialised is high. To estimate how large an error both this conversion and the $L_X-T$ scatter may have introduced, the same method was applied to the data of \citet{1997ApJ...482L..13M}. They measured the temperatures of 38 clusters between a redshift 0.143 and 0.541 using ASCA. A typical error of 20\% was found between the measured temperatures and those calculated with the $L_X-T$ relation, which leads to only a 10\% error in $R_V$. This error, and the previously mentioned uncertainties seem acceptable considering the particular advantages this method has in the removal of the fairly arbitrary decision on the position of the cluster edge and its independence on the heavily fluctuating field density. The calculated values of L$_{bol}$, T, and $R_V$ are show in Table \ref{tab:Virialrad}.

Another advantage of X-ray data is revealed when determining the cluster center. The centroid of the X-ray emission yields the deepest point in the cluster potential well. This often coincides with a bright elliptical galaxy. When this occurred, the center of this galaxy was used as the cluster center. Otherwise, the X-ray centroid was used. When only optical imaging data is present, the cluster center is often determined by the position of the brightest cluster galaxy. It has been demonstrated \citep{2000cofl.work..146L} that this can often lead to an incorrect choice, potentially affecting the measurement of cluster properties.

\subsection{Statistical Field Correction} 
\label{sec:fcorr}

In order to study the cluster galaxy population it was first necessary to correct for the presence of foreground and background field galaxies along the cluster line of sight. A field region was defined for each cluster as the area outside a radius of 3 Mpc. Having a large field of view allowed a field sample to be constructed for each cluster, which should better take into account variations in large-scale structure along the different lines of sight. 

Following the method of \citet{2002MNRAS.331..333P} the probability of each galaxy being a cluster member was determined in a color-magnitude grid of size 0.5 in magnitude and 0.1 in color by comparing the numbers of galaxies in the cluster and field regions. One hundred realisations of the cluster membership were then determined using this probability grid.

\subsection{Fitting the Color-Magnitude Relation}
\label{sec:cmfit}

\begin{figure}
\includegraphics[angle=-90,scale=0.5]{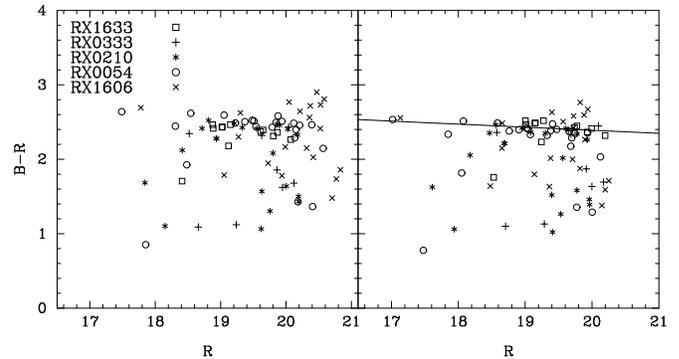}
\caption{\label{fig:cmcomp}$B-R$ Color-Magnitude relations for the 5 lowest L$_X$ clusters before (left) and after (right) K-corrections. The biweight fit to the k-corrected composite is shown as the solid line.}
\end{figure}

\begin{figure*}
\plotone{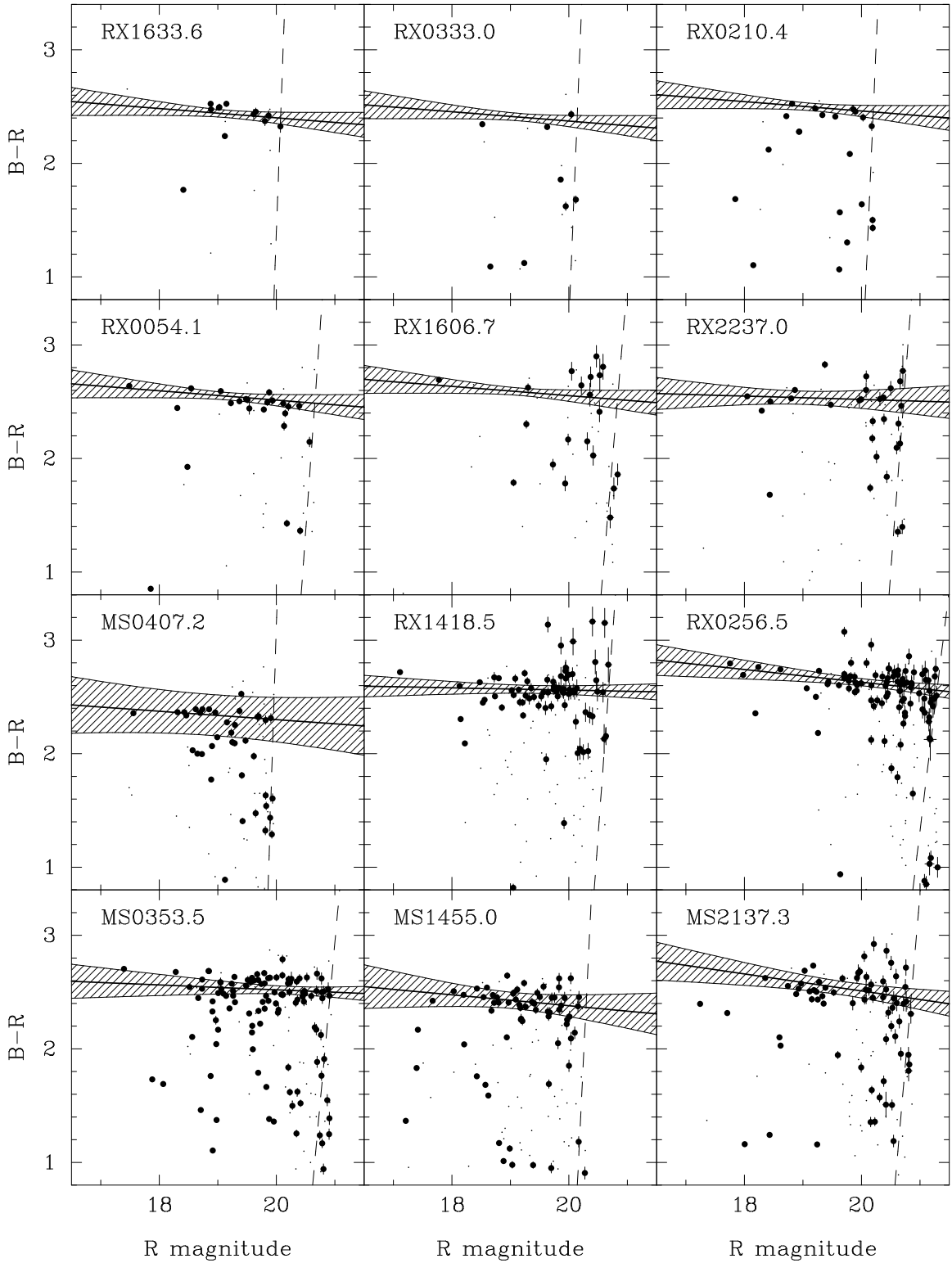}
\caption{\label{fig:cmlzBR} $B-R$ Color-Magnitude relations for all 12 clusters. All galaxies to half a virial radius and to a limiting magnitude of M$_{Vrest}$ = -20 are shown. The large points are galaxies designated as cluster members and the small points as field members in one iteration of the statistical background subtraction. The solid near-horizontal line shows the median biweight fit to the relation and the shaded area the 3 sigma error on that fit from the background correction. For the five lowest L$_X$ clusters the fit shown is that to the composite cluster k-corrected back to the individual cluster redshift. The dashed vertical line shows the color dependent magnitude cut  (see Section \ref{sec:calcfb}).}
\end{figure*}

Although the color-magnitude distribution is dominated by the red galaxies defining the relation to be fit, there is still significant contamination from bluer galaxies. To overcome this contamination, a robust biweight fit was made to the data, which assigns increasingly lower weights to outlying points. \citet{1990AJ....100...32B} give scale (S$_{BI}$) and location estimators (C$_{BI}$) based on the biweight method. Using these estimators, the slope (M) and normalization (C) of the CMR are determined in turn. The CMR residual was defined as $res = (M * magnitude + C) - color$. First $res$ was minimized in S$_{BI}$ with C = 0. This yielded the slope of the relation M. Then $res$ was minimized in C$_{BI}$ with M fixed, giving the normalization of the relation. The procedure was carried out for all clusters in each realization of the background subtraction, in both colors, within a radius of R$_V$/2, to a limiting magnitude M$_V$ = -20. The final fit for each cluster was then determined by taking the median fit of the 100 realizations, and the error given as the $rms$ of these fits.

For the richer clusters, good fits were obtained to their strong red sequences. However, in some of the poorer clusters, the sparsely populated red-sequences were not well defined and hence the fits were poor. To improve on this, a composite of the five lowest L$_X$ clusters (L$_X$ $<$ 10$^{44}$ ergs s$^{-1}$) was made. The fit to this composite CMR would then be used for all of these clusters individually. To construct the composite, it was first necessary to k-correct the galaxy colors and magnitudes to account for the differences in redshift, and hence filter bandpass, between the individual clusters. This was done using	 the K-corrections of \citet{1995PASP..107..945F} with mean corrections of amplitude 0.08, 0.06 and 0.03 in $B$, $V$ and $R$ respectively. Figure \ref{fig:cmcomp} shows 1 of the 100 composite color-magnitude plots before and after K-corrections were applied, with the resultant biweight fit.

\label{sec:CMRLx}
\begin{figure*}
\plottwo{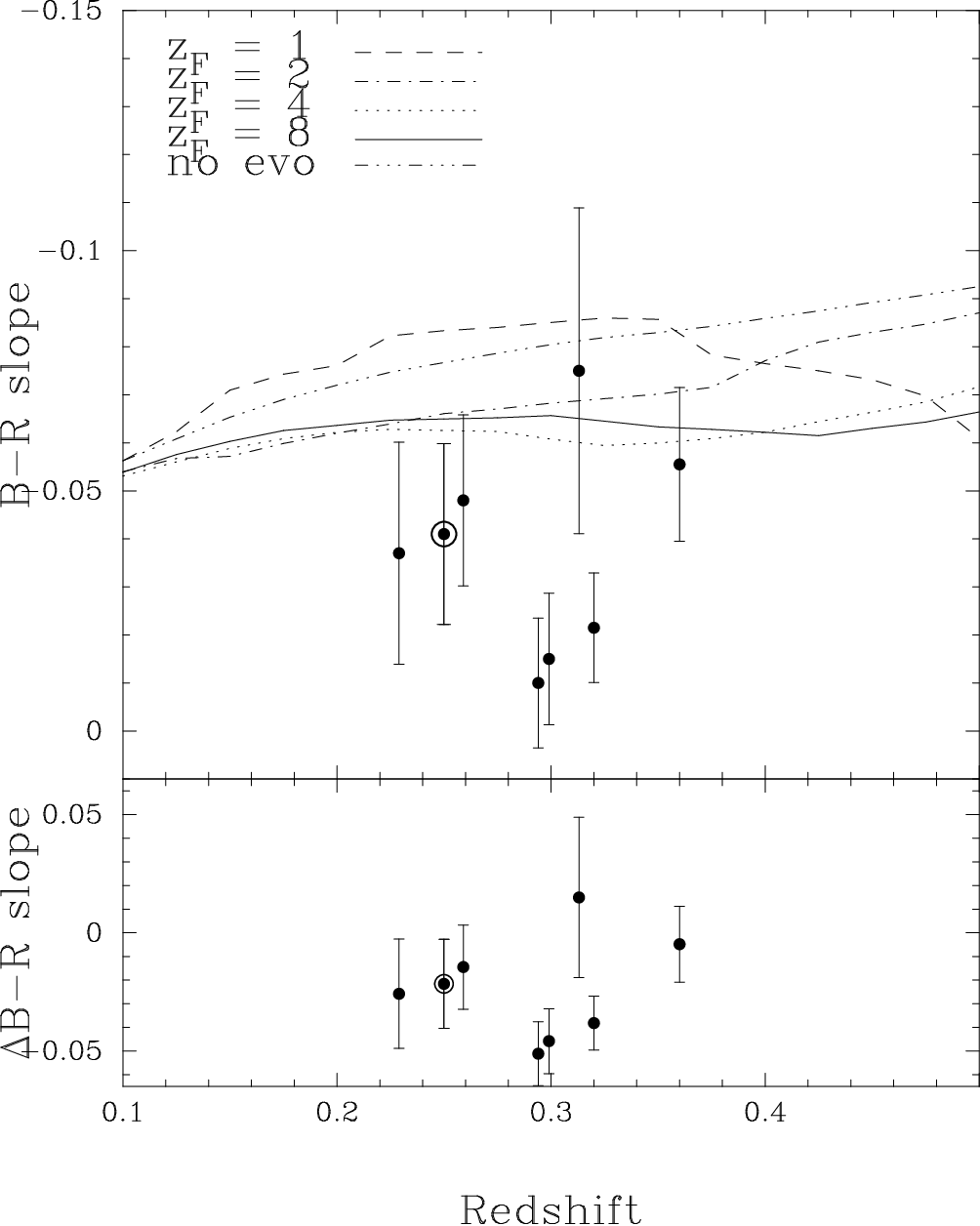}{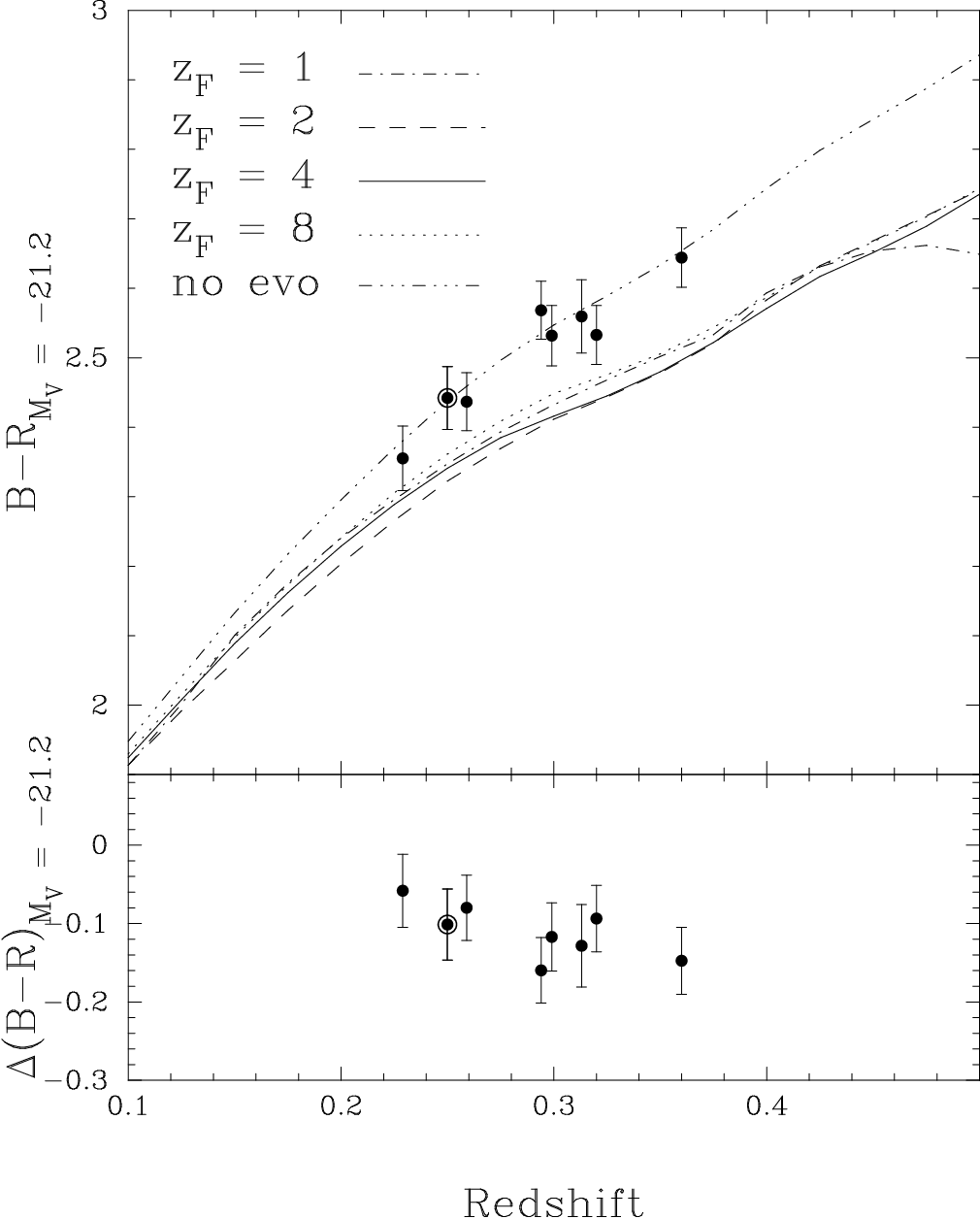}
\caption{\label{fig:cmzBR}The slope (left) and zero-point (right) of the $B-R$ CMR as a function of redshift. The lines show the model predictions for different formation redshifts (see text for details). The bottom plots show the difference between the z$_F$ = 4 model and the measured slopes. The 5 lowest L$_X$ clusters fitted as a composite are shown as one point (the dot inside a circle) at their mean redshift.}
\end{figure*}

\begin{table}
  \begin{center}
    \caption{$B-R$ CMR fit parameters and completeness.\label{tab:cmfitB-R}}
    \begin{tabular}{c  c  c  c  c} 
      \multicolumn{1}{c}{Cluster } &
      \multicolumn{1}{c}{Slope} &
      \multicolumn{1}{c}{Zero-Point }&
      \multicolumn{1}{c}{Target Mag} &
      \multicolumn{1}{c}{Comp}\\
      \multicolumn{1}{c}{} &
      \multicolumn{1}{c}{} &
      \multicolumn{1}{c}{} &
      \multicolumn{1}{c}{M$_V$ = -19} &
      \multicolumn{1}{c}{$R$-Band}\\
      \hline \hline 
      RX1633.6 & -0.041 $\pm$ 0.018 & 3.17 $\pm$ 0.35 & 21.08 & 23.6\\
      RX0333.0 & -0.041 $\pm$ 0.018 & 3.19 $\pm$ 0.35 & 21.15 & 23.7\\
      RX0210.4 & -0.041 $\pm$ 0.018 & 3.28 $\pm$ 0.35 & 21.20 & 23.5\\
      RX0054.0 & -0.041 $\pm$ 0.018 & 3.33 $\pm$ 0.35 & 21.67 & 23.4\\
      RX1606.7 & -0.041 $\pm$ 0.018 & 3.37 $\pm$ 0.35 & 21.84 & 23.4\\
      RX2237.0 & -0.015 $\pm$ 0.014 & 2.82 $\pm$ 0.26 & 21.73 & 23.3\\
      MS0407.2 & -0.037 $\pm$ 0.023 & 3.04 $\pm$ 0.43 & 20.96 & 22.9\\
      RX1418.5 & -0.010 $\pm$ 0.013 & 2.76 $\pm$ 0.26 & 21.68 & 23.3\\
      RX0256.5 & -0.056 $\pm$ 0.016 & 3.74 $\pm$ 0.32 & 22.31 & 23.7\\
      MS0353.6 & -0.022 $\pm$ 0.011 & 2.95 $\pm$ 0.22 & 21.94 & 23.3\\
      MS1455.0 & -0.048 $\pm$ 0.017 & 3.34 $\pm$ 0.33 & 21.30 & 22.7\\
      MS2137.3 & -0.075 $\pm$ 0.033 & 4.01 $\pm$ 0.67 & 21.87 & 23.3\\
      \hline
    \end{tabular}
  \end{center}
    {The median $B-R$ CMR slope and zero-points calculated from the biweight fits to the 100 realizations of the background subtraction with the errors given as the $rms$ of these fits and the $R$-band completeness limit estimated from the NMR.}
\end{table}

Figure \ref{fig:cmlzBR} shows $B-R$ versus $R$ color-magnitude diagrams respectively for the 12 clusters. This filter combination was chosen because it approximates $U-V$ versus m$_V$ at rest, thus spanning the 4000 $\AA$ break and providing a sensitive measure of the age of an old stellar population \citep{1987AJ.....94..899D,1983ApJ...273..105B}. The filled circles show galaxies selected as being in the cluster, and the dots show those in the field for this statistical realization. In all plots, the median biweight fits of all the statistical realizations are shown as solid lines along with the 3 sigma errors on the fits as the shaded regions. 
Details of these fits are given in Table \ref{tab:cmfitB-R}. For the low L$_X$ clusters fitted as a composite, the zero-points have been k-corrected back to the redshift of the individual clusters. We note that fits were also made to CMRs generated with the other filter combinations and that adding these to the analysis below does not change our results.

By inspecting the color-magnitude plots, it can be seen that most of the clusters show clear CMRs. Those that do not, such as RX0333.0 and RX1633.6, are of low richness as might be expected from their very low X-ray luminosities, and thus have very few galaxies with which to define a relation. When these clusters are combined together (Figure \ref{fig:cmcomp}), a relation does become apparent. On plotting the fit from this composite cluster  on the individual color-magnitude diagrams, a concentration of galaxies about the fit is then visible.

\subsection{The Dependence of the Color-Magnitude Relation on Cluster Mass}

 As discussed in Section \ref{sec:intro}, the uniformity of color-magnitude relation provides tight constraints on any variation in the star formation history of early-type galaxies. This section will investigate whether or not the CMR is universal over the large range in cluster X-ray luminosity, and hence mass, present in the sample of clusters studied here.

The properties of the CMR that will be considered here are the slope and zero-point, represented as the color at a fiducial magnitude. In order to investigate how these properties vary with cluster L$_X$, it is necessary to first correct for any dependence on redshift. Figure \ref{fig:cmzBR} shows the slope and zero-point as a function of redshift for the clusters. The low L$_X$ clusters, whose CMRs were fitted together in a composite, are shown as one point in these plots. Also shown on these plots are model predictions for the CMR using the PEGASE2 stellar population synthesis models \citep{1997A&A...326..950F} normalised to the Coma CMR at a redshift 0.023 \citep{1992MNRAS.254..601B}. Clearly there is some discrepancy between the models and the fits is evident. Considering first the evolution of the slope, a large scatter is immediately noticeable, as well as systematically shallower slopes than the models predict. One possibility for the steeper model slopes is the large apertures used to determine the colors. The Coma data, used to normalize the models, used a much smaller aperture to calculate the colors. Since there are radial color gradients in elliptical galaxies, with redder colors nearer the centers, this has the effect of increasing the slope as only the inner regions are sampled in the brightest galaxies. Figure \ref{fig:cmzBR} also shows the zero-point of the CMR of each cluster, defined as the color of a galaxy lying on the fitted CMR that would have an absolute V band magnitude = -21.2 at a redshift zero. Again, there is a discrepancy between the models and the fitted CMRs, with the cluster fits appearing redder than the models. The photometric checks discussed in Section \ref{sec:photchk} suggest that the photometry is much more accurate than the 0.1 - 0.2 magnitude offset visible here. Allied to the consistent offset over both independent data sets, it appears that there may be a problem with the model $B$ magnitudes. This may not be too surprising, \citet{2001AJ....122.2267E} found that the PEGASE models were unable to reproduce the color evolution of Luminous Red Galaxies in the SDSS, particularly in $g-r$, predicting either too red colors at low redshift or too blue at intermediate redshift. We have tied the models to the colors of the low z red galaxies in the Coma cluster which has made the models too blue at the intermediate redshift of the galaxies studied here.

\begin{figure}
\plotone{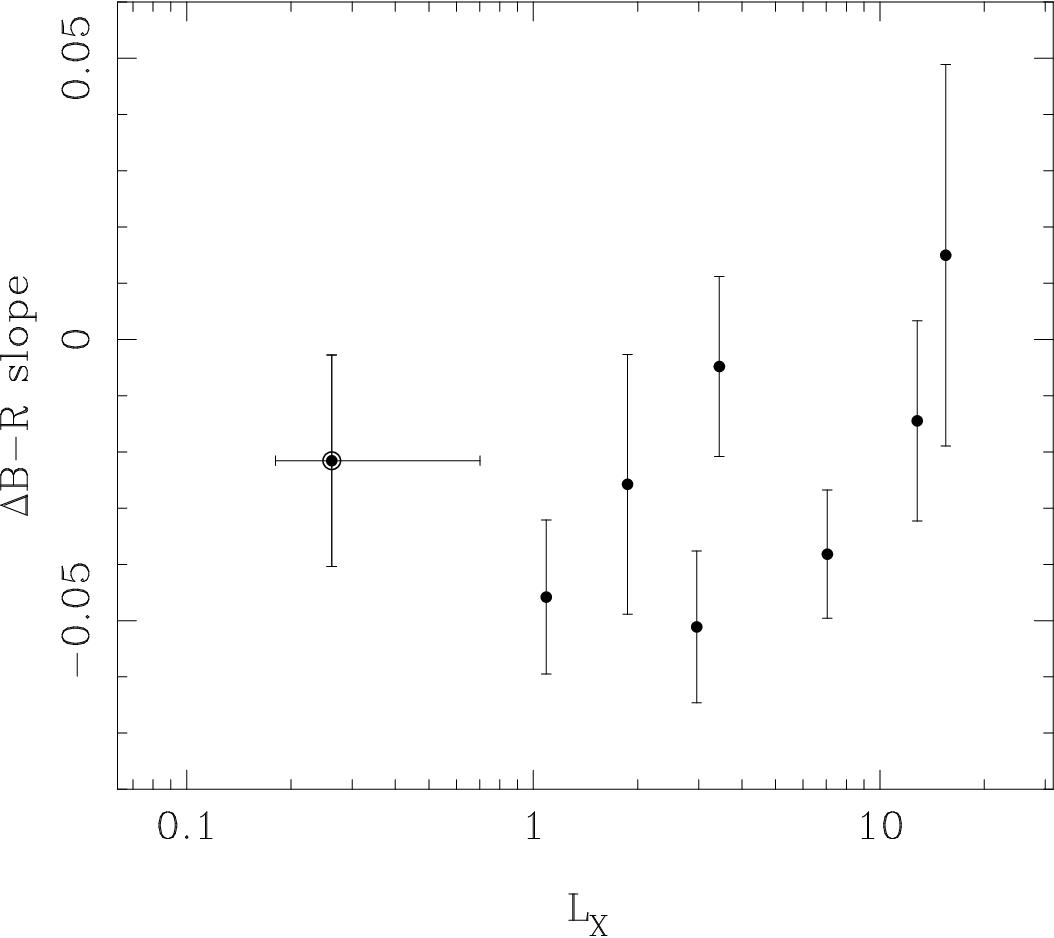}
\caption{\label{fig:cmslopeLx} The difference between the measured CMR and the model CMR slopes as a function of X-ray luminosity. L$_X$ is given for the ROSAT band (0.4 - 2.0 keV) in units of 10$^{44}$ erg s$^{-1}$ .}
\end{figure}

Although the model slopes are offset from the measured slopes, the evolution with redshift appears to be consistent in form. This is seen in the lower left plots in Figure \ref{fig:cmzBR} where no gradient with redshift is apparent. Therefore, it is still possible to use the models to remove the redshift evolution of the slope in order to investigate any dependence on L$_X$, and this is plotted in Figure \ref{fig:cmslopeLx}.

For the zero-points, the models are clearly inadequate. Unlike the slopes, a tight relationship between zero-point and redshift is visible. This enables an empirical determination of the relationship to be made and subtracted in order to investigate the L$_X$ dependence. It is safe to do this, as there is no dependence of L$_X$ on redshift within this sample.

\begin{figure}
\plotone{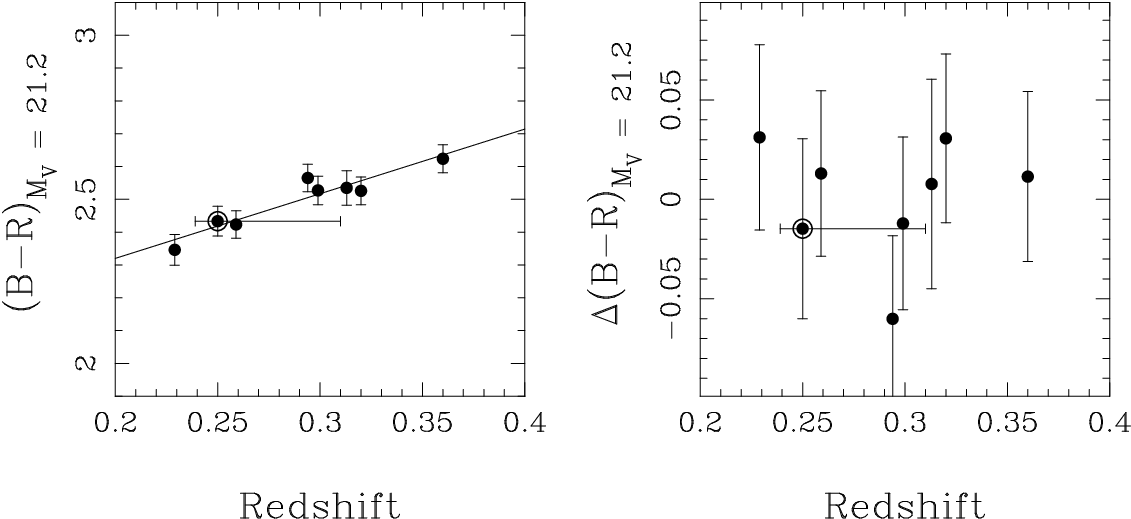}
\caption{\label{fig:cmfidz}The CMR zero-point as a function of redshift (left) and the residual to the fits (right).}
\end{figure}

\begin{figure}
\plotone{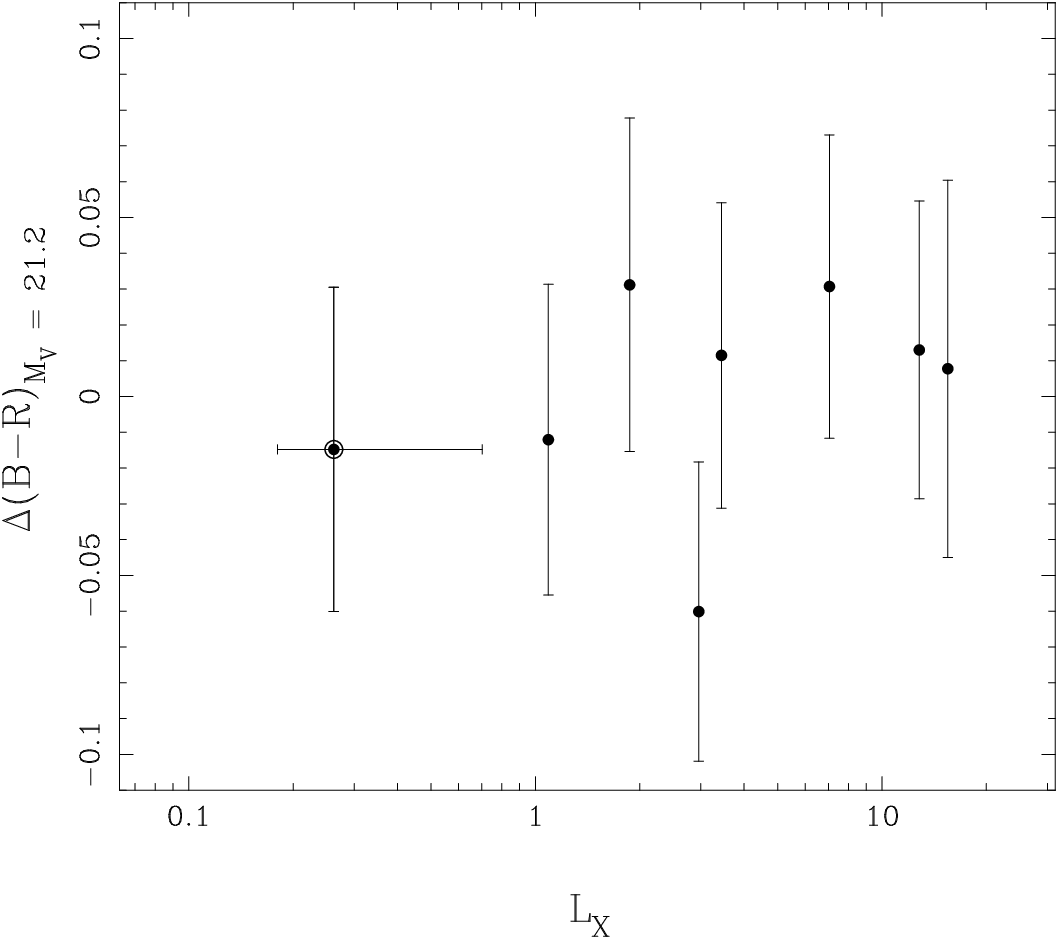}
\caption{\label{fig:cmfidLx}The difference between the measured and the model CMR zero-points as a function of X-ray luminosity (given for the ROSAT band (0.4 - 2.0 keV) in 10$^{44}$ ergs s$^{-1}$).}
\end{figure}

Figure \ref{fig:cmfidz} shows the fits and residuals to the zero-point-redshift relations. The fits are simple, linear least-squares fits and provide good representations of the data. 
Figure \ref{fig:cmfidLx} shows the residuals to the fits as a function of X-ray luminosity.

Inspecting Figures \ref{fig:cmslopeLx} and \ref{fig:cmfidLx}, the color-magnitude relation shows no dependence on cluster X-ray luminosity either in its slope or zero-point over a factor of 100 in X-ray luminosity. This is confirmed by Spearman Rank tests, with correlation probabilities of 82$\%$ and 61$\%$ for the slope and zero-point respectively, consistent with no trend. The scatter in both the slope and zero-point L$_X$ relations is consistent with being driven by the measurement errors. 

Even though the range in X-ray luminosity is large, the number of clusters considered is quite small, and the errors and scatter in these plots quite large. It is therefore worth considering how much of a trend would still be consistent with these plots. To calculate this, a simple least-squares fit weighted by the errors was made to the relations in Figures \ref{fig:cmslopeLx} and \ref{fig:cmfidLx}. A maximum variation of 0.091 in the slope and 0.090 in the zero-point was allowed by the 2 sigma error to the fits over the entire L$_X$ range of the clusters. The change in the slopes allowed is quite large, however, the maximum zero-point change is well constrained.


If the change in zero-point is considered to be due to an age change, the maximum age change allowed can then be estimated. Using the PEGASE2 models, an age change of 1 Gyr corresponds to a color change of $\sim$ 0.05 in $B-R$ at z $\sim$ 0.3. This means that an age difference of greater than $\sim$ 2 Gyrs is ruled out at the 2 sigma level. It therefore appears that the global environment does not have a strong affect on the ages of early-type galaxies in clusters.

\subsection{The Dependence of the Color-Magnitude Relation on Cluster Radius}
\begin{figure}
\plotone{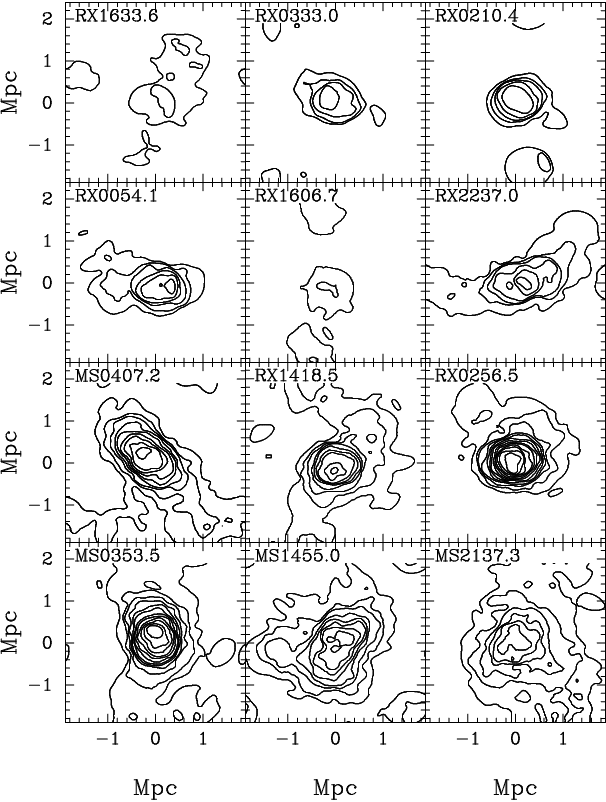}
\caption{\label{fig:surfden}The galaxy surface density distributions. Only galaxies with a color within 0.08 of both the fitted $B-V$ and $V-R$ color magnitude relations are considered. The positions are smoothed with a top hat filter of 500 Kpc in size. The lowest contour represents 3 Mpc$^{-1}$ and each contour increases by 3 Mpc$^{-1}$.}
\end{figure}

In the previous section we investigated how the CMR depends on the global environment of a galaxy cluster, as measured by its X-ray luminosity. Another environmental change we can probe is the variation of local galaxy density as measured by cluster radius. One approach to investigating the CMR as a function of radius would be to try and fit the CMR in radial bins for each cluster. However, rapidly decreasing cluster galaxy densities, and thus increasing field subtraction errors, make this unfeasible. Instead, all the clusters are combined together using a method similar to that described in \citet{2002MNRAS.331..333P}. 
Since we have already shown that the properties of red sequence galaxies are independent of the X-ray luminosity of the cluster in which they reside we can be confident that no significant biases due to the large range in L$_X$ have been introduced when the entire cluster sample is combined together. The lowest L$_X$ clusters, while contributing many fewer galaxies than the highest L$_X$ clusters, do make a significant contribution, with the 5 lowest L$_X$ clusters together adding a similar number of cluster galaxies as one of the richest. Figure \ref{fig:surfden} shows the surface density of the galaxies lying close to the CMR for each cluster. The clusters are generally centrally concentrated with approximately circular density contours, and any irregularities should be smoothed out when the clusters are combined.

Before combining, the CMR fits were subtracted from the galaxy colors of each cluster, $i.e.$ for each galaxy the color difference ($\Delta(B-R)$) from the fitted CMR was calculated.
This was repeated for every galaxy in the cluster area (within 2 $R_V$) and for every galaxy in a combined field sample. The combined field sample was constructed by including all the galaxies in the cluster fields at a radius greater than  2.5 $R_V$ from the cluster center. A combined sample is used here, rather then a field sample for each cluster as in the CMR fitting, both to allow a larger radius from the cluster center to be probed, and to maintain sufficient numbers to minimize the statistical errors in the field subtraction.

For each cluster, the cluster and field samples were cut to a magnitude limit equivalent to a V absolute magnitude of -20 at the cluster redshift, $\sim$M$_*$ + 2. The cluster galaxies were then split into annuli of width one quarter of a virial radius, and binned into histograms of $\Delta$color with bins of 0.1 magnitude. The field sample was similarly binned into a histogram. For each radial annulus, the field histogram was scaled according to the ratio of the areas and then subtracted from the cluster histogram. This is a much simpler method of field correction than that used when fitting the CMR, as the magnitude dependence has been removed, thus improving the number statistics in each bin, and enabling a more reliable background subtraction over the same size color bin.
The final step was to sum all the field subtracted histograms in each annulus to construct a composite of all the clusters. 

When combining a number of clusters, it is important to ensure the same sampling has been made for each cluster, both spatially and photometrically. The use of the virial radius ensured that the spatial scales of clusters were matched to each other. In order to compare the colors of galaxies at different redshifts, a K-correction is normally applied. This method was used when calculating the magnitude limit for each cluster using the K-corrections of \citet{1995PASP..107..945F}. However, by calculating the color difference from the fitted CMR, the equivalent of a k and evolutionary correction has been applied based on the empirical colors of the galaxies lying on the CMR of each cluster. This `correction' is likely to be more accurate than any determined by models, as it has been determined using the data on which it was to be applied. Its only limitation is that the equivalent of one correction was applied to all galaxies regardless of how close their color lies to that of the CMR. However, as the galaxies of interest here are those lying close to the CMR, and the redshift range is small, any systematic errors would be much smaller than those introduced by using model K-corrections.

\begin{figure*}
\plotone{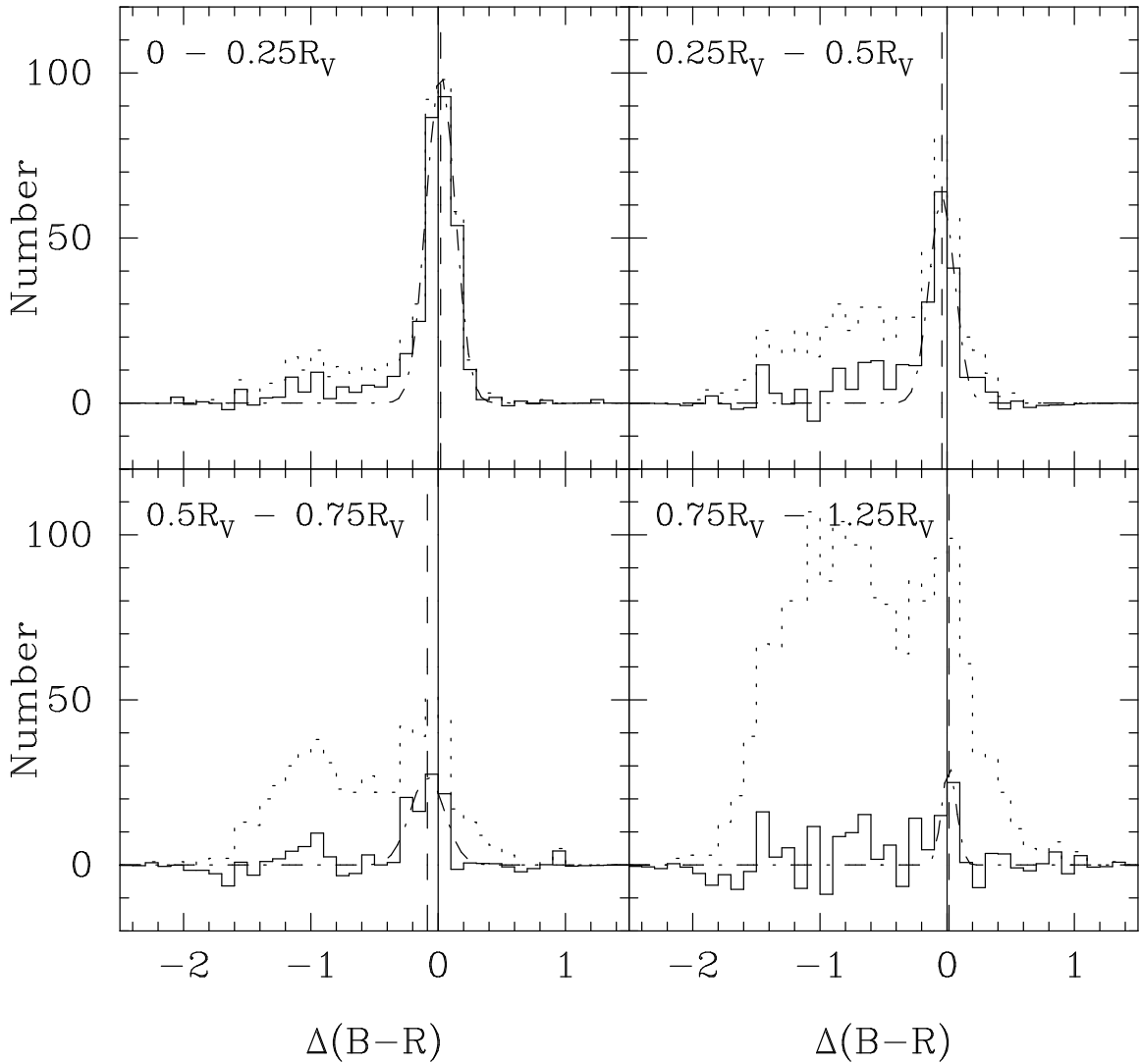}
\caption{\label{fig:cmradB-R}$\Delta(B-R)$ color histograms as a function of radius for the combined clusters. The solid histogram is field subtracted, and the dotted histogram includes the field galaxies. The dot-dash line shows the fitted Gaussian, and the dashed line shows the position of the Gaussian peak. }
\end{figure*}

Figure \ref{fig:cmradB-R} shows the combined histograms, in 4 radial bins, for the 12 clusters. The solid histogram shows the field corrected data, and the dotted histogram shows all data prior to the field correction. Each panel shows the histogram in the radial bins of 0.25 $R_V$, except the final panel which shows the histogram in a larger bin, between 0.75 and 1.25 $R_V$, in an attempt to increase the signal caused by the low cluster density at these radii. The central bin (top left panel) shows the clear peak of the cluster CMR. This gets smaller with increasing radii and disappears in the final bin (bottom right panel) at $\sim$ $R_V$. 

\begin{table}
  \begin{center}
    \caption{Gaussian fit parameters\label{tab:gaussfit}}
    \begin{tabular}{c  c  c} 
      \multicolumn{1}{c}{Radius} &
      \multicolumn{1}{c}{Peak} &
      \multicolumn{1}{c}{Sigma}\\
      \hline \hline
	0 - 0.25$R_V$       &  0.020 $\pm$ 0.009 & 0.159 $\pm$ 0.012\\
	0.25$R_V$ - 0.5$R_V$  & -0.041 $\pm$ 0.012 & 0.142 $\pm$ 0.024\\
	0.5$R_V$ - 0.75$R_V$  & -0.083 $\pm$ 0.035 & 0.184 $\pm$ 0.041\\
	0.75$R_V$ - 1.25$R_V$ & -0.026 $\pm$ 0.133 & 0.139 $\pm$ 0.108\\

      \hline
    \end{tabular}
  \end{center}
\end{table}

To investigate any change in the color of the CMR with cluster radius, we wished to determine if and how the cluster peak color and width change. To do this, a Gaussian was fitted iteratively to the histogram data around the peak between -0.4 and 0.4 in $\Delta(B-R)$, thus ignoring the very bluest galaxies which become more significant at larger radii. The results of these fits are given in Table \ref{tab:gaussfit}, and are plotted on  Figure \ref{fig:cmradB-R} as the dot-dashed curves, with the center of the Gaussian fit shown as a dashed line. To generate errors on the fits, random Gaussian noise scaled to each bin's Poisson error was added to each histogram bin, and then the fit was recalculated. This was repeated 10000 times with the final fits and errors calculated as the mean and $rms$.
A Gaussian fit was used to define the peak instead of other location estimators such as the biweight, as it could be fitted to data including negative bins, whereas the biweight cannot.

The peak of the Gaussian fits show a clear trend of becoming bluer with increasing radius in the first 3 bins, with the fourth bin being very noisy with a large error on the fit. A total shift bluer of 0.10 $\pm$ 0.036 magnitudes was observed over 0.75 $R_V$. The fact that the first bin shows a positive peak is reassuring, since the subtracted CMR was fitted within 0.5 $R_V$ and so would lie between the first two radial bins considered here.
The width of the CMR shows no evidence of any significant change with radius. However, the errors on the width are large, so definite conclusions regarding the CMR width are difficult.

A clear variation in the color of the CMR with radius is seen for a combined sample of 12 intermediate redshift clusters. Removing the slope of the CMR before the radial comparison is made ensures that any color change is independent of the magnitude distribution of the galaxies in a radial bin. This is important, as the magnitude distribution will change, with the most massive, brightest and hence reddest galaxies being located in the cluster center. \citet{2002MNRAS.331..333P} measured the change in $B-R$ color as a function of radius in a combined sample of 11 rich X-ray selected clusters at a redshift of $\sim$ 0.1. Since their clusters all had similar X-ray luminosities, they did not normalize the individual radii to a universal scale, but just summed the clusters, measuring the radial color change in Mpc$^{-1}$. To directly compare the two trends, the virial radius of a cluster with the average X-ray luminosity and redshift of their sample was calculated using the same method as for the clusters presented here, and the radii scaled accordingly.

\begin{figure}
\plotone{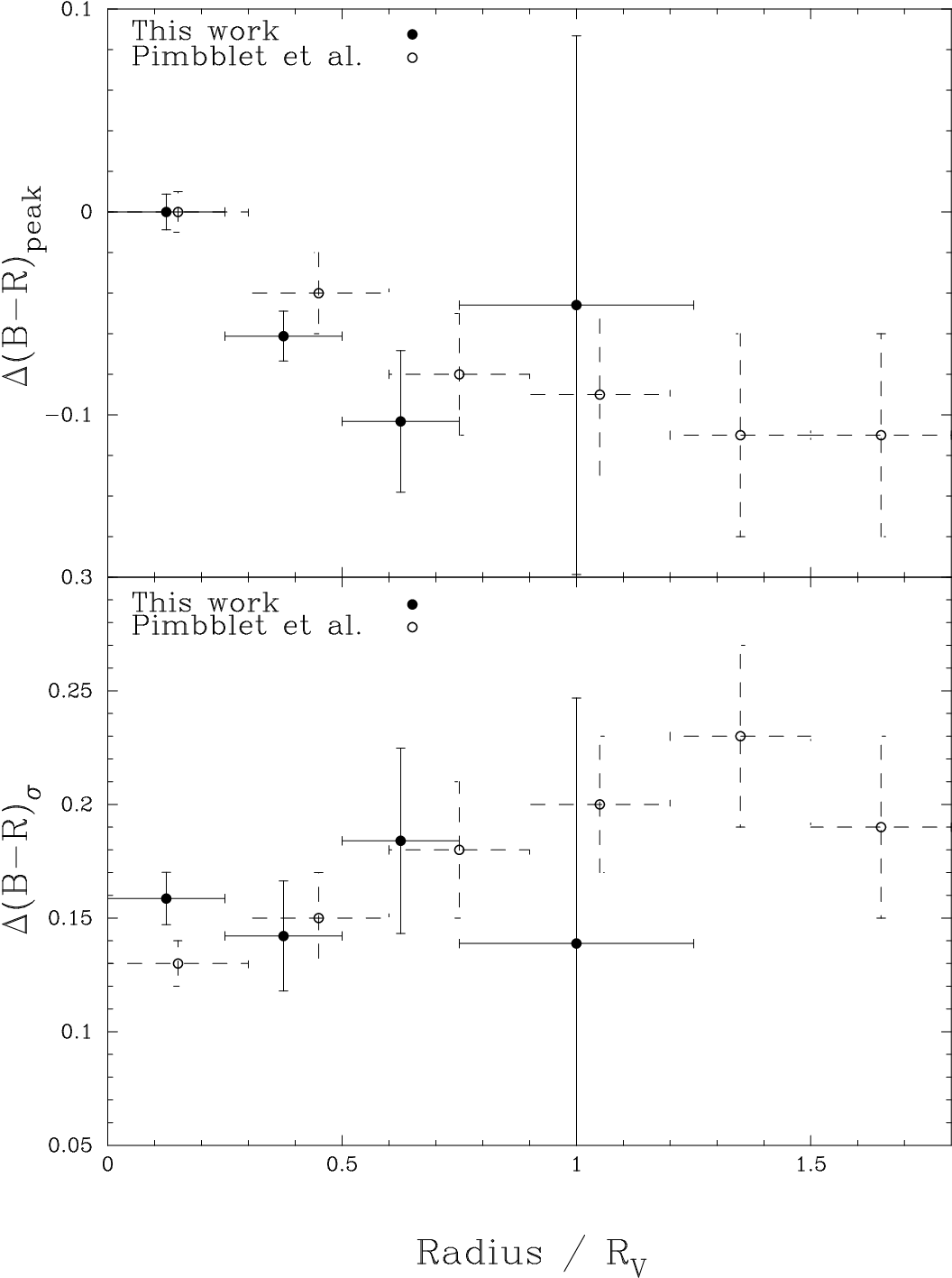}
\caption{\label{fig:cmpeaksigpim} The CMR peak position (top) and width (bottom) as a function of virial radius for both this work (filled circles) and that of \citet{2002MNRAS.331..333P} (open circles). The peak position is shown relative to the central value.}
\end{figure}

A comparison of the peak and width of the fitted Gaussians of the two studies is shown in Figure \ref{fig:cmpeaksigpim}. Considering firstly the change in peak position, the two trends are similar, with the one presented here being slightly steeper than \citet{2002MNRAS.331..333P}. Assuming that the color change reflects a change in the age of the galaxies on the CMR \citep{2002MNRAS.331..333P,2001MNRAS.326.1547T,1996ApJ...471..694A}, a steeper trend would be expected at a higher redshift due to the younger overall age of the galaxies, and thus the larger fractional age gradient with radius. Turning to the width of the fitted Gaussian, there is little evidence of any trend in the data presented here, contrary to that found by \citet{2002MNRAS.331..333P}, although the large errors make definite conclusions difficult (see also \citet{2004ApJ...615L.101B}).

Several other authors have reported similar radial color changes of the CMR. \citet{1996ApJ...471..694A} studied a single cluster, A2390, at a redshift of 0.23, with both photometry and spectroscopy, over a wide area. They found that the colors of galaxies on the CMR became bluer with radius with a gradient of $d(g-r)/dlog_{10}(r)$ = -0.079 out to 3.7 Mpc, equivalent to $\sim$ 1.15 $R_V$ calculated from the X-ray temperature measurement of \citet{2001MNRAS.324..877A}. \citet{2001MNRAS.326.1547T} reported a bluening of the $U-V$ color magnitude relation in Coma as a function of cluster radius. Converting to the $g-r$ color used by Abraham et al., they report a change of $d(g-r)/dlog_{10}(r)$ = -0.024 $\pm$ 0.005. \citet{2002MNRAS.331..333P} similarly converted their $B-R$ color to $g-r$ and find $d(g-r)/dlog_{10}(r)$ = -0.061 $\pm$ 0.011, which is intermediate between the other two studies.

Converting the $B-R$ color difference of the cluster sample presented here to $g-r$ produces $d(g-r)/dlog_{10}(r)$ = -0.081 $\pm$ 0.019, consistent with the trend of \citet{1996ApJ...471..694A} at the same redshift.
Using simple evolutionary models presented in section \ref{sec:CMRLx}, the age change corresponding to these color changes is calculated, assuming that the galaxies are formed at a redshift $>$ 2. This enables the rate of age change per log(r) to be determined. The results of these calculations are shown in Table \ref{tab:dage}.

\begin{table}
  \begin{center}
   \caption{\label{tab:dage}}
    \begin{tabular}{l  c  c  c  c} 
      \multicolumn{1}{l}{Study} &
      \multicolumn{1}{c}{z} &
      \multicolumn{1}{c}{$\Delta$(g-r)/log(r)}&
      \multicolumn{1}{c}{$\Delta$(g-r)/Gyr}&
      \multicolumn{1}{c}{$\Delta$(Age)/log(r)}\\
      \hline \hline
	Terlevich et al.  & 0.023 & -0.024 $\pm$ 0.005 & 0.0083 & -2.9 $\pm$ 0.6\\
	Pimbblet et al.  & 0.12  & -0.061 $\pm$ 0.011 & 0.0151 & -2.5 $\pm$ 0.7\\
	Abraham et al.   & 0.23  & -0.079 $\pm$ 0.02 & 0.0283 & -2.8 $\pm$ 0.7\\
	This Work        & 0.28  & -0.081 $\pm$ 0.019& 0.0323 & -2.5 $\pm$ 0.6\\
      \hline
    \end{tabular}
  \end{center}
    {Rate of change of $g-r$ peak color as a function of log projected radius, the calculated change in color per Gyr of an old stellar population and the resulting rate of change of age as a function of projected radius from 4 photometric studies of clusters.}
\end{table}

All four studies produce very consistent results with a luminosity weighted age change of $\sim$ 2.5 Gyrs per log(r), accounting for the radial color change. Both \citet{1996ApJ...471..694A} and \citet{2002MNRAS.331..333P} explain this change as a result of an increase in younger S0 galaxies contaminating the CMR at larger radii. \citep{1997ApJ...490..577D} show that the fraction of S0s in clusters declines with increasing redshift suggesting that S0 galaxies in clusters have been formed relatively recently. They further suggest that infalling spiral galaxies in the field are transformed into S0 galaxies by the truncation of their star-formation. This may result in a gradient of younger S0s with cluster radius. \citet{2002MNRAS.331..333P} suggest that the increasing width of the CMR histogram they observe supports this, with a larger range of galaxy colors expected from the spread in age of the recently formed S0s. Unfortunately due to the large errors it is not possible to distinguish between no trend and that of \citet{2002MNRAS.331..333P} here. 
Recently \citet{2004ApJ...601L..29H} have conducted a similar study of bulge dominated galaxies in the SDSS. They find evidence of a small shift in the peak color ($<$ 0.02 mag $g-r$) and a broadening of the blue wing of galaxy color distribution with density further supporting this hypothesis.

\subsection{The Blue Fraction}
\label{sec:calcfb}

In the previous sections we have investigated the properties of the red cluster galaxies as a function of cluster environment. We now turn to the blue cluster galaxy population and calculate the fraction of blue galaxies (f$_b$) following the method of \citet{1984ApJ...285..426B}.
Butcher \& Oemler defined the blue galaxies as those 0.2 magnitudes bluer than the CMR in rest frame $B-V$. One approach to calculating f$_b$ would therefore be to k-correct all of the data to the rest frame $B-V$. However, a simpler approach suggested by \citet{2001MNRAS.321...18K} is to k-correct the blue galaxy criteria to the cluster redshift. A $B-V$ color 0.2 mag bluer than an elliptical galaxy falls between the colors of an Sab and Sbc at a redshift of zero. The blue fraction cut was thus defined as the color at the same position between a Sab and a Sbc at the cluster redshift, using the colors and K-corrections of \citet{1995PASP..107..945F}. The blue fraction cuts are listed in Table \ref{tab:fb}. Butcher and Oemler also defined f$_b$ to be measured to a limiting M$_{V}$ of -20 at z = 0. At higher redshifts, this limiting magnitude depends on spectral type. Again, the \citet{1995PASP..107..945F} colors and K-corrections were used to determine the limiting magnitude for each cluster, which are shown as the dot-dot-dot-dash lines in Figure \ref{fig:cmlzBR}. These lines are close to vertical in most cases, due to the clusters being chosen at redshifts where the measurement filters map approximately onto rest UBV. The limiting magnitude for each cluster at the color of an elliptical galaxy is given in Table \ref{tab:fb}.
 
\begin{table}
  \begin{center}
     \caption{Blue fractions.\label{tab:fb}}
   \begin{tabular}{l  c  c  c} 
      \multicolumn{1}{l}{Cluster} &
      \multicolumn{1}{c}{Lim m$_R$} &
      \multicolumn{1}{c}{f$_b$ cut} &
      \multicolumn{1}{c}{f$_b$}\\ 
      \hline \hline 
      RX1633.6 &20.076 & 0.182 &   0.06 $^{+0.39(0.29)}_{-0.29(0.15)}$ \\
      RX0333.0 &20.145 & 0.188 &  -0.13 $^{+0.63(0.53)}_{-0.44(0.26)}$ \\
      RX0210.4 &20.201 & 0.072 &   0.21 $^{+0.22(0.20)}_{-0.16(0.13)}$ \\
      RX0054.0 &20.667 & 0.093 &   0.06 $^{+0.19(0.17)}_{-0.14(0.09)}$ \\
      RX1606.7 &20.839 & 0.255 &  -0.01 $^{+0.25(0.20)}_{-0.20(0.13)}$ \\
      RX2237.0 &20.729 & 0.240 &   0.06 $^{+0.24(0.18)}_{-0.20(0.13)}$ \\
      MS0407.2 &19.959 & 0.172 &   0.08 $^{+0.13(0.11)}_{-0.11(0.08)}$ \\
      RX1418.5 &20.678 & 0.234 &   0.10 $^{+0.11(0.08)}_{-0.10(0.06)}$ \\
      RX0256.5 &21.314 & 0.303 &   0.02 $^{+0.08(0.06)}_{-0.07(0.04)}$ \\
      MS0353.6 &20.937 & 0.269 &   0.04 $^{+0.09(0.07)}_{-0.07(0.05)}$ \\
      MS1455.0 &20.300 & 0.200 &   0.07 $^{+0.19(0.12)}_{-0.17(0.09)}$ \\
      MS2137.3 &20.869 & 0.259 &   0.02 $^{+0.13(0.10)}_{-0.11(0.07)}$ \\
      \hline \\
    \end{tabular}
  \end{center}
    {Blue fractions for all clusters. The first error on each f$_b$ is that calculated using the $rms$ background variation, and the second assuming a Poisson error in the background.  }
\end{table}

On calculating the blue fraction, a field correction was again required. However, since f$_b$ is a simple measure which splits all the galaxies into just 2 groups, the complicated grid method was not necessary. Again, the background was determined for each cluster at a radii $>$ 3 Mpc. The galaxies in both the cluster and field areas were then split into blue and red galaxies in the same manner. These numbers were then scaled by the relative corrected areas, which are calculated in the same manner as in Section \ref{sec:cmfit}. The field numbers were then subtracted from the cluster numbers to give f$_b$ simply as residual blue galaxies divided by the residual total galaxy population. The error in f$_b$ was determined using the Poisson errors on the numbers of galaxies in the blue and red, cluster and field samples using the equations of \citet{1986ApJ...303..336G}.





When determining the field errors, it was important to consider the clustering of galaxies. To estimate this effect, the variance in the field region on the size of the cluster area was determined for each image. This was done by splitting the field into a grid, with each grid square having the same area as the cluster within which f$_b$ is measured. The number of blue and red galaxies were calculated in the manner described above, and the mean and $rms$ of this distribution determined. The $rms$ was then used as the error in the field numbers as it represents how the field beneath the cluster is typically varying, based on the local galaxy distribution. This $rms$ error was always larger than the Poisson error on the whole background by a factor of $\sim$ 8, resulting in a increased error in f$_b$ of $\sim$ 50\%.

\begin{figure}
\plotone{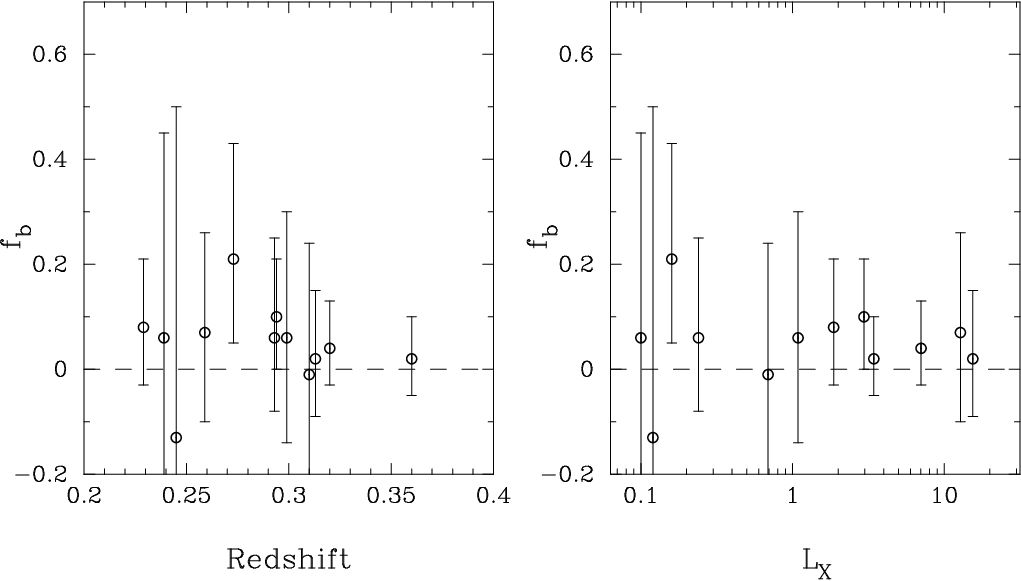}
\caption{\label{fig:fbzlx} Blue fraction as a function of redshift (left) and X-ray luminosity in units of 10$^{44}$ ergs s$^{-1}$ (right). Error bars are calculated using the $rms$ variation in the background.}
\end{figure}

In order to investigate any dependence of the blue fraction on redshift and X-ray luminosity in a manner consistent with Butcher and Oemler, f$_b$ was first calculated within a radius of $R_{V}$/3 approximating R$_{30}$.
Table \ref{tab:fb} shows the resultant values of f$_b$ in observed $V-R$ ($\sim B-V$ at rest). Errors are given for purely Poisson variations on the number counts and for the $rms$ background variations. 
Figure \ref{fig:fbzlx} shows this data plotted as a function of both redshift and X-ray luminosity. It is clear that no obvious correlations exist and that there is a large scatter. Spearman Rank correlation tests confirm this. The large scatter and small range in redshift mean that it would be very difficult to draw any conclusions on the evolution of f$_b$ with redshift from this data alone. 

 

This is not the case for the lack of a correlation of f$_b$ with L$_X$. This cluster sample covers a factor of $\sim$ 100 in X-ray luminosity with an even distribution of clusters throughout this range. The lack of any trend suggests that the blue fraction is independent of the global properties of the cluster, and that the distribution of galaxy types is the same in all cluster cores regardless of mass. This is in agreement with the work of \citet{2002MNRAS.337..256B}, who show that the distributions of star formation rates in low L$_X$ clusters at z $\sim$ 0.25 are indistinguishable from those of X-ray luminous clusters at similar redshifts, and \citet{2004MNRAS.351..125D} who show that f$_b$ is independent of the velocity dispersion of local clusters in the 2dF Galaxy Redshift Survey. However, \citet{2001ApJ...548L.143M} and \citet{2003PASJ...55..739G} both find a trend of increasing f$_b$ with decreasing cluster richness. It should be noted that both studies use a fixed physical aperture within which they measure f$_b$. This would correspond to a larger fraction of the virial radius, or R$_{30}$, for the poor clusters, and could be driving this effect.

\begin{figure}
\plotone{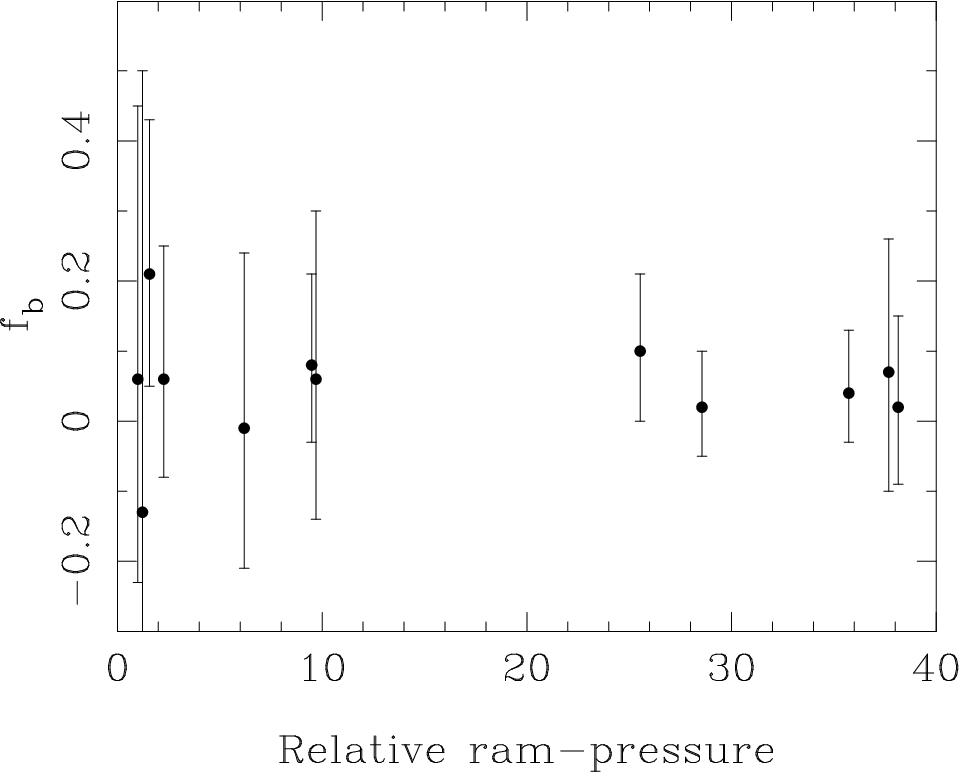}
\caption{\label{fig:fbram}The blue fraction as a function of relative average ram-pressure. The ram-pressure is normalized to give the lowest L$_X$ cluster RX1633.6 a value of 1.}
\end{figure}

It is worth briefly considering the implications of an independence of f$_b$ from L$_X$ on the transformation mechanisms discussed in Section \ref{sec:intro}. The most likely explanation for the Butcher-Oemler effect is that as time goes on there is a declining rate of accretion of field galaxies on to clusters, combined with a declining star formation rate in these field galaxies, along with a mechanism to truncate star formation within the cluster environment \citep{2001MNRAS.321...18K}. If that truncation mechanism depends on the cluster mass then we might expect to see a correlation between f$_b$ and L$_X$. For instance, if the truncation mechanism becomes ineffective for some mass threshold, all the accreted star forming galaxies would remain blue resulting in a higher blue fraction. For ram-pressure stripping we are able to directly determine the ram-pressure efficiency from the cluster L$_X$.
Figure \ref{fig:fbram} shows the rest frame $B-V$ blue fraction as a function of relative ram-pressure for all the clusters. The relative ram-pressure is calculated using the relations $P_{ram} =  \rho_{ICM} V_{gal}^2$ and $V_{gal} \propto kT_{CL}^{0.55}$ and $\rho_{ICM}^2 \propto L_X / kT_{CL}^2$, where $\rho_{ICM}$ is the density of the intracluster medium, $V_{gal}$ is the mean velocity of a galaxy through the cluster, and $T_{CL}$ and L$_X$ are intracluster medium temperature and X-ray luminosity respectively. The ram-pressure is normalized to give the lowest L$_X$ cluster (RX0210.4) a value of 1. Figure \ref{fig:fbram} shows that the average ram-pressure varies by a factor of almost 40 in this sample, and yet no dependence with f$_b$ is observed. Further detailed numerical simulations \citep{1999MNRAS.310..663S} and more sophisticated analytical models \citep{1999ApJ...516..619F} show that galaxies in the lowest L$_X$ clusters observed here would not be expected to experience any ram-pressure stripping, even in the very central regions of the clusters where both their velocities and the cluster gas density are highest. Therefore, it seems unlikely that ram-pressure stripping, or other mechanisms which depend on the global cluster properties such as tidal stripping or harassment, are the sole mechanism determining the blue fraction.  

\begin{figure}
\plotone{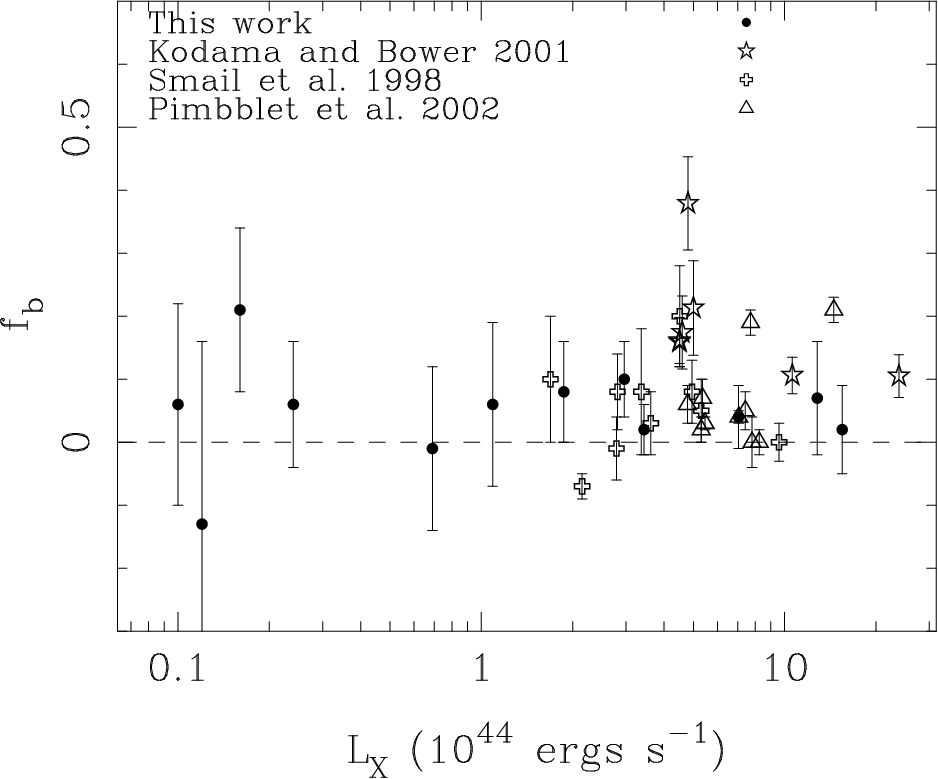}
\caption{\label{fig:fblitLx} Blue fraction as a function of L$_X$, for 4 X-ray selected cluster samples of \citet{2001MNRAS.321...18K}, \citet{1998MNRAS.293..124S}, \citet{2002MNRAS.331..333P} and this work. The blue fractions from this work are calculated within R$_V$/3.}
\end{figure}

Figure \ref{fig:fblitLx} shows the rest $B-V$ blue fractions along with 3 X-ray selected cluster samples \citep{2001MNRAS.321...18K,1998MNRAS.293..124S,2002MNRAS.331..333P}. This illustrates the extension in the range of L$_X$ sampled by this study, and demonstrates the lack of a correlation between f$_b$ and L$_X$ over all these studies. 

A number of authors have reported that most clusters show increasing blue fractions with radius \citep{1984ApJ...285..426B,1996ApJ...471..694A,2002MNRAS.330..755F,2001MNRAS.321...18K}. This is consistent with the morphology-density relation \citep{1994ApJ...430..107D,1980ApJ...236..351D} and the observed increase in the average star-formation rate with cluster radius \citep{1999ApJ...527...54B,1996ApJ...471..694A,2003ApJ...584..210G,2002MNRAS.334..673L}.

\begin{figure}
\plotone{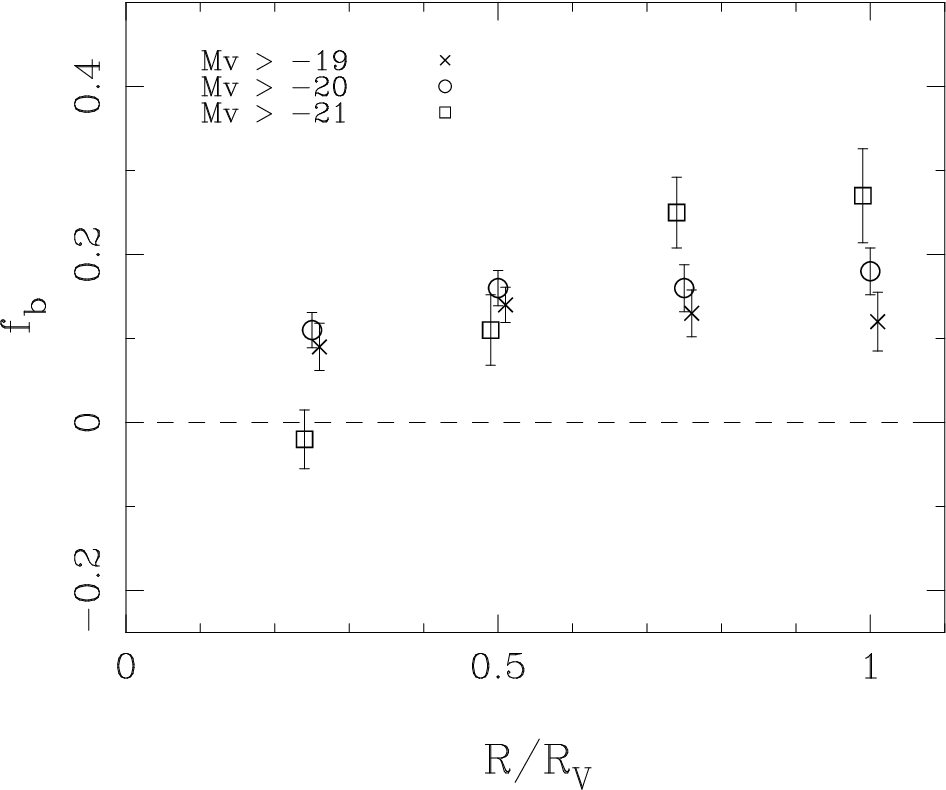}
\caption{\label{fig:fbrcom} Blue fraction as a function of virial radius for the combined cluster sample. The blue fraction is calculated to three magnitude limits in M$_V$, -19 (crosses), -20 (circles) and -21 (squares)}
\end{figure}

Figure \ref{fig:fbrcom} shows the average blue fraction calculated within a series of radii for the whole cluster sample. Values of f$_b$ within each radius are also calculated for 3 magnitude limits: the standard BO limit, and 1 magnitude either side of this. The trend of increasing f$_b$ with radius is confirmed in this cluster sample with the main increase seeming to occur within R$_V$/2, with a near constant fraction beyond. The brightest galaxies show the largest trend, but the errors preclude making any quantitative conclusion. This is not surprising, as the cluster core is expected to be dominated by bright elliptical galaxies.

\section{Conclusions}

We have analyzed the colors of bright galaxies in a sample of 12 X-ray selected clusters of galaxies at z $\sim$ 0.3. The clusters cover a large range in X-ray luminosity (10$^{43}$ - 10$^{45}$ erg s$^{-1}$) and hence mass, with the lowest L$_X$ clusters representing some of the least massive studied at this redshift. Further more, the wide field of our imaging has allowed us to study the galaxy colors out to the virial radius of each cluster as determined by its X-ray luminosity, well beyond the core region typical of previous studies at this redshift.

We have shown that the red galaxies in our clusters form a tight color-magnitude relation and measure its slope and zero-point. After removing the redshift evolution, we found that neither the slope nor zero-point depend strongly on cluster X-ray luminosity and hence mass. However, these constraints were limited by both the small number of clusters and the accuracy of the CMR fitting caused by the statistical field correction required in purely photometric studied of galaxy clusters. By combining all the clusters together, we were able to measure the change in the zero-point of the CMR as a function of cluster radius. We found that the CMR became bluer by 0.1 $\pm$ 0.036 in $B-R$ out to 0.75 R$_V$. If this was driven purely by an age change, it corresponds to a difference of $\sim$ 2.5 Gyrs.    

We also measure the fraction of blue star forming galaxies as defined by Butcher and Oemler in our clusters. Again we find that there is no dependence of f$_b$ on cluster X-ray luminosity and hence mass. However, we do again find a dependence on radius, particularly for the brightest galaxies.

Overall, we can find no dependence of cluster galaxy population on the global cluster environment, $i.e.$ cluster mass, but we do find a dependence on local environment, $i.e.$ galaxy density. The lack of any trends with cluster mass suggest that mechanisms like ram-pressure stripping, tidal stripping, or harassment, which depend strongly on cluster mass are unlikely to be solely responsible for changing the galaxy population from the blue star forming galaxies that dominate low density environments to the red passive galaxies that dominate clusters.

\acknowledgments
We thank the referee, Ian Smail, for his thorough reading of this paper and insightful comments. We also thank David Pinfield for the use of his defringing code and Mariangela Bernardi, Chris Miller, Phil James, Andrew Newsam and John Porter for useful discussions. 
DAW acknowledges the receipt of PPARC a studentship and NSF grant XXXX. DJB acknowledges the support of NASA contract NAS8-39073.

\clearpage

\end{document}